\documentclass[aps,prd,twocolumn]{revtex4}
\usepackage{amsmath,amssymb,epsfig,bbm,tmbib}
\usepackage{color}
\newcommand{\comment}[1]{{}}
\def\beq{\begin{equation}}
\def\eeq{\end{equation}}
\def\beqn{\begin{eqnarray}}
\def\eeqn{\end{eqnarray}}

\def\cl{C_{\ell}}

\def\l{\left}
\def\r{\right}

\def\2gcm{\textrm{g cm$^{-2}$}}

\def\modu#1{\l |{#1}\r |}
\def\av#1{\l \langle{#1}\r \rangle}
\def\hmpc{\:{h}^{-1}\mathrm{Mpc}}

\def\H0{\ensuremath{\mathrm{H}_0}}

\def\nn{\nonumber}

\newcommand{\bmm}[1]{{\mathbf{#1}}}

\newcommand{\bm}[1]{\ensuremath{\mbox{\boldmath $#1$}}}
\def\bx{\bmm{x}}

\renewcommand\vec[1]{\mathbf{#1}}
\def\be{\begin{equation}}
\def\ee{\end{equation}}
\def\bea{\begin{eqnarray}}
\def\eea{\end{eqnarray}}
\def\bk{\bmm{k}}
\def\mse{\mathrm{MSE}}
\def\btheta{\bm{\theta} }
\def\bell{\bm \ell }
\def\nres{N_{res}}
\def\ntap{N_{tap}}

\def\eqd{\,{\buildrel \mathbf{d} \over =}\,}
\begin{document}
\title{Efficient Power Spectrum Estimation for High Resolution CMB Maps}
\author{Sudeep Das}
\email{sudeep@astro.princeton.edu}
\author{Amir Hajian}
\email{ahajian@princeton.edu}
\author{David N. Spergel}
\email{dns@astro.princeton.edu}
\affiliation{Department of Astrophysical Sciences, Princeton University, Princeton, NJ 08544.}
\date{\today}
\begin{abstract}
Estimation of the angular power spectrum of the Cosmic Microwave Background (CMB) on a small patch of sky is usually plagued by serious spectral leakage, specially when the map has a hard edge. Even on a full sky map, point source masks can alias power from large scales to small scales producing excess variance at high multipoles. We describe a new fast, simple and local method for estimation of power spectra on small patches of the sky that minimizes spectral leakage and reduces the variance of the spectral estimate. For example, when compared with the standard uniform sampling approach on a $8$ degree $\times$ $8$ degree patch of the sky with $2\%$ area masked due to point sources, our estimator halves the errorbars at $\ell=2000$ and achieves a more than fourfold reduction in  errorbars at $\ell=3500$. Thus, a properly analyzed experiment will have errorbars at $\ell=3500$ equivalent to those of an experiment analyzed with the now standard technique with $\sim 16-25$ times the integration time.
\end{abstract}
\maketitle
\section{Introduction}
Cosmic Microwave Background (CMB) is a statistically isotropic \citep{HS06}  and Gaussian \citep{Komatsu} random field. If we ignore secondary effects, all of the information in high resolution CMB maps is encoded in the angular correlation function or equivalently, in the angular power spectrum, $\cl$. The angular power spectrum is widely used to estimate the cosmological parameters.  Accurate measurements of  angular power spectrum are needed for precise estimation of cosmological parameters.\par
Over the past decade, CMB power spectrum has been measured over a large range of multipoles, $\ell$,  by various groups \citep{COBE, Nolta, Hinshaw, ACBAR, QUAD}. And more experiments are under way to measure the $C_\ell$ on smaller scales with high accuracy \citep{ACT, SPT, Planck}. Most power spectrum analyses use uniform or noise weighted maps. This performs reasonably well for power spectra that have nearly equal power in equal logarithmic intervals of multipoles,  {\it i.e.} $\ell \leq 1000$ for the CMB. For smaller scales, (larger $\ell$) this method is non-optimal, as we show in section \ref{master}. CMB power spectrum estimated from an incomplete sky map is the underlying full-sky  power spectrum convolved with the power spectrum of the mask. This leads to coupling of modes in the estimated power spectrum. For high resolution experiments such as ACT and SPT which will map the small scale  anisotropies of the CMB on small patches of the sky, this mode-mode coupling will be a serious problem. The reason is that CMB power spectrum is very red on those scales (it falls off as $\ell^{-4}$ at large $\ell$) and hence is highly vulnerable to the leakage of power due to mode-mode coupling. There are two methods to remedy this:  to taper the map near the sharp edges, and to pre-whiten the CMB power spectrum. In order to minimize the loss of information due to applying a taper to the map, we use the multitaper method \citep{Percival+93}. This method involves weighting the map with a set of orthonormal functions which are space limited but maximally concentrated in the frequency domain.  Power spectrum of each of these tapered maps is a measurement of the power spectrum of that map with a different amount of mode coupling. Final power spectrum is obtained from a particular linear combination of these tapered power spectra that minimizes the bias in the estimated power spectrum. The use of multiple tapers also reduces the error-bars in the measured power spectrum. \par
Mode coupling is less harmful if the map has a nearly white power spectrum.  Traditionally, an inverse covariance matrix weighting is used in analysis to prewhiten the maps \citep{Tegmark}. This method works well, but is a computationally expensive, non-local operation and may be complicated to implement, specially for  high resolution experiments \citep{Dore, Smith}.  We propose a simple and local prewhitening operator in real space (\S~\ref{prewhiten}) that is fast to implement and reduces the bias due to the leakage of power. This method prevents the unnecessarily large error bars at $\ell \gtrsim 1500$ due to the point source masks. Usually masks  have sharp edges and holes at the positions of point sources. This leads to a mode-coupled power spectrum that is highly biased at large $\ell$.  Deconvolution of the mode-coupled power spectrum is a well-studied problem in the CMB data analysis literature \citep{2002:Hivon} and has been applied to many experiments. But deconvolution of a highly biased power spectrum leads to large error bars in the final power spectrum at large $\ell$. The mode coupling problem will be worse for the upcoming set of CMB experiments as bright point sources will be much more of a limiting foreground at high resolution.\par
As we show in \S~\ref{master}, prewhitening followed by the multitaper method for power spectrum estimation reduces the error-bars (specially at large $\ell$) in the decoupled power spectrum (cf. Figs. \ref{errorCompare} \&~\ref{PWMasteredErrors}).\par
We begin with a review of the multitaper method in one-dimension in \S~\ref{review} and discuss the salient features of the method, generalizing it to the two-dimensional case. Next, we discuss the statistical properties of multitaper spectrum estimators. As a simple application,  we demonstrate the method in context of CMB power spectrum estimation in \S~\ref{application}. Next, we formulate the prewhitening method (\S~\ref{prewhiten}) and apply it to the case of CMB power spectrum estimation in presence of masks. In \S~\ref{master}, we describe the algorithm for deconvolving the power spectrum and the implications of the multitaper method and prewhitening in its context. We summarize and conclude in \S~\ref{conclusion}.
\section{A Brief Review of the Multitaper Method \label{review}}
The problem of estimating the power spectrum of a stationary, ergodic process, sampled at discrete intervals and observed over a finite segment of its duration of occurrence, is an old and well-studied one (for an extensive treatise, see \cite{Percival+93}). Several methods have been traditionally used for power spectrum estimation in one-dimension. These include non-parametric methods like the periodogram, the lag-window estimators, Welch's overlapped segment averaging \citep{1161901} and the Multitaper  method \citep{Thomson:1982p3137}, and parametric methods like the maximum likelihood estimation. In this paper, we generalize the one-dimensional Multitaper method to two-dimensions and adapt it to handle real data with noise and masks on a two-dimensional flat Euclidean patch. We discuss its applications specifically in the case of CMB power spectrum estimation. \par
The most basic spectral estimation method is to take the square of the Fourier Transform (FT) of the observed data. Taking the FT of a finite segment of data is equivalent to convolving the underlying power spectrum with the power spectrum of a top-hat function. As the latter has substantial sidelobe power, it leads to spectra leakage and the resulting spectrum is highly biased.
Most of the non-parametric methods for power spectrum estimation utilize some kind of a data taper (a smooth function that goes smoothly to zero at the edges of the observed segment) to minimize the effect of spectral leakage. Such smoothing reduces the bias in the estimator at the cost of lower spectral resolution. As the taper down-weights a fraction of the data, one is left with an effectively lower sample size. Since tapering also smooths in frequency space, it essentially leads to a loss of information which is reflected in the increased variance of the final estimate. The first attempt at ameliorating these disadvantages of using a data taper was addressed in a seminal paper by \cite{Thomson:1982p3137} which laid down the basis of the Multitaper Method (MTM). The basic idea of MTM is to apply multiple orthogonal tapers with optimal spectral concentration to minimize the loss of information due to tapering. 
\subsection{Notations}
Throughout this paper, we will refer to spatial coordinates as the $\bx$ space (this may be an angular coordinate in radians on the sky, or a comoving distance in $\hmpc$, etc.) and the reciprocal space as the $\bk$ space (which would be the multipole space $\ell$, or the Fourier modes in $\hmpc$, etc).  The continuous Fourier Transform conventions adopted here are,
\beqn
\tilde F(\bk)&=&\int d^{n}\bx  ~ F(\bx) ~ \exp(-i\bk\cdot\bx)\:\:\: \text{(forward)}\\
F(\bx)&=&\int \frac{d^{n}\bk}{(2\pi)^{n}} ~ \tilde F(\bk) ~ \exp(i\bk\cdot\bx) \:\: \text{(inverse)},
\eeqn
where $n$ is the dimensionality of the space. \par
For a stationary process, $F(\bx)$ the power spectrum is defined as, 
\beq
(2\pi)^{n} P(k)=\av{\tilde F^*(\bk)\tilde F(\bk)} .
\eeq

\subsection{1-D Multitaper Theory}
Although power spectrum estimation for the CMB is an inherently two-dimensional problem, we will begin by discussing the multitaper theory in one dimension. This is because the essential features  of the theory are easier to understand in one dimension and can be trivially generalized to higher dimensions. \par
 We consider a stationary, stochastic, zero mean process $F(x)$ sampled at $N$ discrete points, $x_j$ sampled at regular intervals of size $\Delta x$. Let $P(k)$ be the true underlying power spectrum of the process. Our problem is to estimate $P(k)$ using the sample of size $N$.\par
 The Nyquist frequency for the problem is  given by $f_N=k_N/(2\pi)=1/(2\Delta x)$ and the fundamental frequency by $f_0=k_0/(2\pi)=1/(N\Delta x)$. In the following, we will assume $\Delta x=1$ for simplicity.\par
Let us contemplate windowing our data by some function $G(x)$, generating the product, 
\beq
y(x_j)=G(x_j) F(x_j)
\eeq
and taking the power spectrum of the windowed data as,
\beq
(2\pi)\hat P(k)\equiv \tilde y^*(k) \tilde y(k) = {\modu{\sum_{j=1}^N  G(x_j) F(x_j) e^{-i k x}}}^2,
\eeq
where $\tilde y$ is the Fourier transform of $y(x)$. The quantity $\hat P$ can be thought of as an estimator of $P(k)$, such that its ensemble average is related to $P(k)$ through
\beq
 \av{\hat P(k)} = \int_{-k_N}^{k_N} \frac{d k'}{(2\pi)}  {\Gamma}(k-k') P(k').
\eeq
This means that $\hat P$ is an estimator of the true power spectrum convolved with a \emph{spectral window function},
\beq
{\Gamma}(k)=\modu{\tilde G(k)}^2.
\eeq
If ${\Gamma}$ could be designed such that ${\Gamma}(k) = (2\pi) \delta(k)$ then $\hat P$ would be an exact unbiased estimator of $P$. However, a function like $G$ which is spatially limited in extent cannot be arbitrarily concentrated in the frequency space. If the window function is a top-hat, its power spectrum will be a $ \mathrm{sinc}^2$ function\footnote{ $ \mathrm{sinc}$ refers to the \emph{sinus cardinus} i.e. $\sin{x}/x$. Power spectrum of a two-dimensional top-hat window is a product of two $ \mathrm{sinc}^2$ functions. } with substantial sidelobes. This will lead to the aliasing of power on various scales, an effect known as spectral leakage or mode coupling. Mode coupling is specially damaging for a spectrum which is highly colored or structured. \par
The multitaper method (MTM) consists of finding a set of orthogonal window functions or tapers, which are maximally concentrated in some predetermined frequency interval. With the set of tapers, one can generate several approximately uncorrelated estimates of the power spectrum. This is superior to the plain Fourier Transform (Periodogram) because it not only attempts at remedying mode coupling errors but also helps decrease the uncertainty in the estimated power spectrum by generating independent realizations of the same power spectrum with information from different section of the data. We formulate the method below.\par
We desire a set of  tapers, such that each of them is spatially limited,  $G_j$ $(j=1,...,N)$  and has its power ${\Gamma}(k)$  optimally concentrated in some frequency interval, $k\in [-2\pi W,2\pi W]$, with $W<f_N \equiv k_{N}/(2\pi)$. Here we have introduced the shorthand notation $G_{j}$ for $G(x_{j})$. Concentration is quantified by the following quantity, 
\beq
\beta^2(W)=\frac{\int_{-2\pi W}^{2\pi W} {\Gamma}(k) dk}{\int_{-k_N}^{k_N} {\Gamma}(k) dk }.
\eeq
which is basically the fractional power of the taper inside the desired interval. Remembering that, 
\beq
\tilde G_k=\sum_{j=1}^{N} G_j e^{-i k x_j},
\eeq
the above equation can be re-written as, 
\beq
\beta^2(W)=\sum_{j'=1}^{N}\sum_{j=1}^{N} G^*_j\frac{\sin[2\pi W(j-j')]}{\pi (j-j')}   G_{j'}\left/\sum_{j=1}^{N} \modu{G_j}^2\right.
\eeq
It is easy to see that the sequence $G_j$ that will maximize $\beta(W)$ must satisfy, 
\beq
\sum_{j'=1}^{N}\frac{\sin[2\pi W(j-j')]}{\pi (j-j')} G_{j'} =\lambda_\alpha (N,W) G_j,
\eeq
for $ j=1,...,N.$ This can be immediately recognized as an eigenvalue problem, 
\beq
\mathbb{A} \vec G=\lambda_\alpha (N,W) \vec G
\eeq
where $\mathbb{A}$ is the $N\times N$ Toeplitz matrix,
\beq
\mathbb{A}_{jj'}=\frac{\sin[2\pi W(j-j')]}{\pi (j-j')}
\eeq
and $\vec G$ is the vector, index limited from $j=1$ to $j=N$ which has the highest concentration in the frequency interval $[-W,W]$. Here $\alpha$ denotes the indices of the different eigenvalues of the problem. The solution to this eigenvalue problem is well known \citep{1978ATTTJ..57.1371S}. There are $N$ nonzero eigenvalues of the problem denoted by $\lambda_{\alpha}$ ($\alpha= 0, 1, ..., (N-1)$) with corresponding eigenvectors $\vec v^{\alpha}$. The elements of each of the $N$ eigenvectors consist of a finite subset of the discrete prolate spheroidal sequence (DPSS). The zeroth eigenvector $\vec{ v}^0(N,W)$ which has the highest eigenvalue $\lambda_0$ is composed of the zeroth order DPSS, the  eigenvector  $\vec{v}^1(N,W)$ having eigenvalue  $\lambda_1<\lambda_0$ is composed of first order DPSS sequence and so on.\par
\begin{figure}[htbp]
\begin{center}
\includegraphics[scale=0.55]{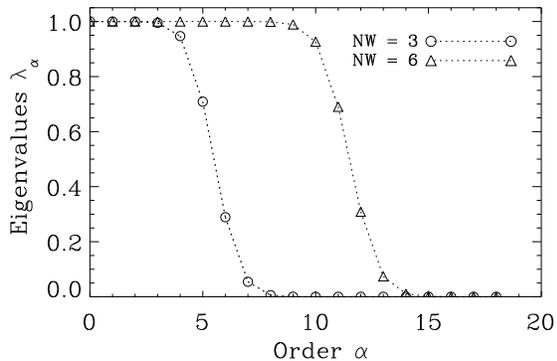}
\caption{Eigenvalues corresponding the different orders of DPSS tapers. Two cases with $NW = 3$ and $6$ are shown for $N=50$. Spectral concentration of the tapers rapidly worsen beyond $\alpha = 2 NW -1$. }
\label{eigenValues}
\end{center}
\end{figure}
\begin{figure*}[t]
\includegraphics[scale=0.6]{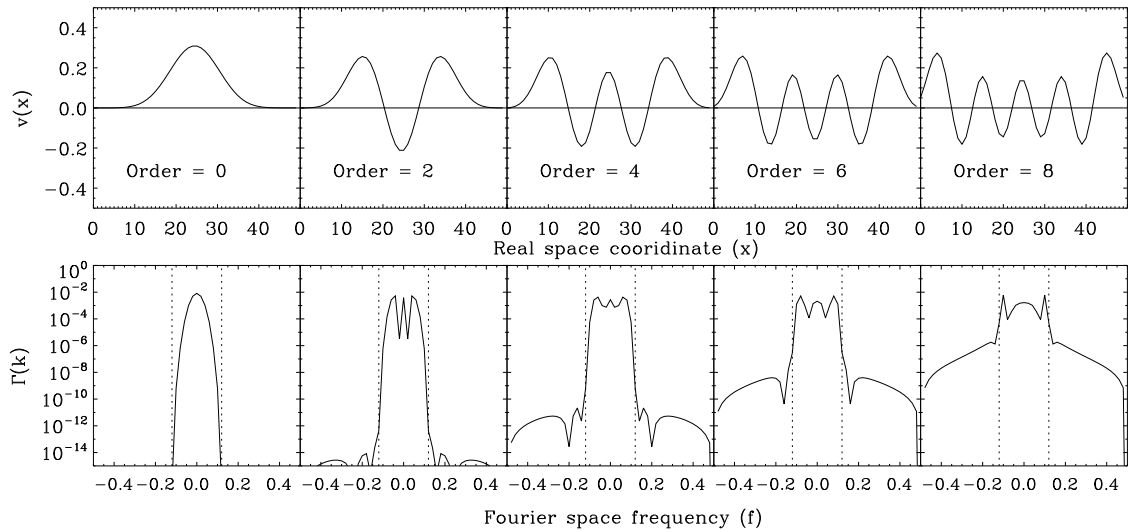}
\caption{Examples of DPSS tapers and the corresponding spectral window functions for the case $N=50$ and $NW=6$. {\emph{Upper panel:}}  Real space form of the tapers of orders  $0$, $4$, $6$ and $8$. {\emph{Lower panel: }} The spectral window functions corresponding to the tapers in the upper panel. For simplicity, the window functions are shown only in  the range $-3 W<f< 3 W$ of frequency. The  vertical dotted lines denote the edges of the bandwidth $(-W,W)$ within which the tapers are designed to be optimally concentrated. Tapers are ordered such that spectral leakage progressively increases for tapers of higher order. \label{tapersAndPowers}}
\end{figure*}

Some salient properties of the $N$ eigenvectors and eigenvalues are as follows,
\begin{enumerate}
\item The eigenvalues are bounded by $0$ and $1$:
$$0 < \vec v^{\alpha} < 1 .$$
\item The eigenvectors are orthogonal and can be standardized so that they are orthonormal,
  $$ {\vec v^\alpha}\cdot {\vec v}^\beta=\delta_{\alpha\beta}$$
\item The eigenvectors form a basis for an $N$-dimensional Euclidean space. 
\item Usually the  eigenvectors are ordered according to decreasing eigenvalues.  The first $2NW-1$ eigenvalues are close to unity (most concentrated) and the eigenvalues rapidly fall to zero thereafter. This behavior is illustrated in Fig.~\ref{eigenValues}. The number $2NW-1$ is often referred to in the MTM literature as the \emph{Shanon Number}. 
\end{enumerate}
One-dimensional DPSS taper generation algorithms are usually included in standard signal processing softwares (e.g. the $\tt dpss$ module of Matlab). For the purpose of this paper, we used a Fortran 90 implementation of the original algorithm by \cite{1993:Bell}.
Examples of DPSS tapers and their corresponding spectral window functions are displayed in Fig~\ref{tapersAndPowers}, where the gradual worsening of the leakage properties of the tapers are apparent.

The Fourier transforms of the tapers,
\beq
\tilde{{v}}^{\alpha}(k) = \sum_{j} v^{\alpha}(x_{j}) e^{-i k x_{j}}
\eeq
 also have the interesting properties:
\begin{enumerate}
\item They are orthonormal over the frequency range $-k_{N}<k<k_{N}$,
$$\int_{-k_{N}}^{k_{N}} \frac{d k}{2\pi} \tilde{{v}}^{\alpha *}(k) \tilde{{v}}^{\beta}(k) = \delta_{\alpha\beta} .$$
\item They are also orthogonal over the frequency domain $-2\pi W <k< 2\pi W$,
\beq
\label{orthoInW}
\int_{-2\pi W}^{2\pi W} \frac{d k}{2\pi} \tilde{{v}}^{\alpha *}(k) \tilde{{v}}^{\beta}(k) = \lambda_{\alpha}\delta_{\alpha\beta} .
\eeq
This means that the functions $\tilde{\vec{v}}^{\alpha}/\sqrt{\lambda_{\alpha}}$ form an orthonormal set on the inner interval $-2\pi W <k< 2\pi W$.
\end{enumerate}

Having generated the {\it{eigentapers}}, we can form $M$ approximately uncorrelated estimators of the power spectrum with the first $M$ eigenvectors having the best concentration,
\beq
(2\pi) \hat P^\alpha(k)={\modu{\sum_{j=1}^N  v^\alpha_j F_j e^{-i k x_{j}}}}^2
\eeq
where $\alpha = 0, ..., M-1$.\par
Then we can form a weighted mean of the tapered power spectra, often called the \emph{eigenspectra}, to generate the final estimate of $P(k)$, the simplest form of which is,
\beq
\label{mtmSpec}
P^{\mathrm{MTM}}(k)=\frac{\sum_{\alpha=0}^{M-1} \lambda_\alpha \hat P^\alpha(k)}{\sum_{\alpha=0}^{M-1} \lambda_\alpha}
\eeq
It can be shown that this is the optimal way of estimating $P$ in the case where the process is white noise. For colored spectra, a more sophisticated approach is required, which leads to the adaptive multitaper method (AMTM) to be discussed in the following section.\par
The important point to note here is that the MTM or variants of it, aim to restore the information lost to a single taper algorithm by weighting different parts of the data by an orthogonal set of tapers, thereby reducing the variance in the final estimate. Lower variance comes at the cost of decreased spectral resolution. One should bear in mind that the choice of the resolution $W$ is completely dependent on the analyst. Remembering that the number of useful tapers is $(2NW-1)$ and a better spectral resolution, i.e. smaller $W$ means that there will be fewer tapers to work with. The choice of $W$ will in most cases be dictated by the type of and  the features in the power spectrum being estimated.
\subsection{Adaptive MTM}
As noted above, if the underlying spectrum is white and only modest spectral resolution is needed in the analysis, many eigenspectra can be simply combined with scalar weights to get a good estimate of the true spectrum. However, this is not the case for spectra which are colored and have large dynamic range. For example, the Cosmic Microwave Background (CMB) power spectrum $\cl$ falls like $\ell^{-4}$ beyond $\ell\sim 1000$. In cases such as this, only the first few tapers are good at avoiding aliasing of power due to mode coupling. As more and more tapers are used, the estimated power spectrum gets more and more biased. \par
The adaptive multitaper method (AMTM) aims at mitigating this problem, thereby allowing the use of a larger number of tapers even for a colored spectrum. In the following we briefly sketch the AMTM method. \par
According to the Cramer spectral representation of a stationary process \citep{1940:Cramer}, a stationary zero mean process can be represented as,
\beq
\label{cramerRep}
F(x) = \int_{-k_{N}}^{k_{N}} e^{ikx}dZ(k)  
\eeq
for all $x$, where $dZ$ is an orthogonal incremental process  \citep{doob:172,1988:Priestly}.  The random orthogonal measure $dZ(k)$ has the properties,
\beq
\av{dZ(k)} = 0 ; \: \: \av{{\modu{dZ(k)}}^{2}} = P(k) dk .
\eeq
where $P(k)$, as before, is the true underlying spectrum.\par
The Fourier transform of the data weighted by an eigentaper can be written as,
\beqn 
\tilde y^{{\alpha}}(k)  &=& \sum_{j=1}^{N}   v^{\alpha}(x_{j}) F(x_{j}) e^{{-ik\;x_{j}}}\\
& =& \int_{-k_{N}}^{k_{N}} \tilde v^{\alpha}(k-k') dZ(k')
\eeqn
using \eqref{cramerRep}. Note that this contains information from the entire Nyquist range.\par
Now consider the case where the signal $F(x)$ is convolved with a perfect bandpass filter from $k-2\pi W$ to $k+2\pi W$ to yield the unobservable yet perfect eigencomponent,
\beq
\tilde{\cal Y}^{\alpha}(k) = \int_{k'=k-2\pi W }^{k+2\pi W} \frac{\tilde v^{\alpha}(k-k')}{\sqrt{\lambda_{k}}} dZ(k')
\eeq
which contains information only from the interval $(-2\pi W, 2\pi W)$. Note that in accordance with \eqref{orthoInW} we have used the correct orthonormal form for FT of the tapers inside this interval.\par 
The quantity $\tilde{\cal Y}(k)$, although fictitious in the sense that it cannot be computed from the data, has the desirable property that,
\beq
\av{{\modu{\tilde{\cal Y}(k)}}^{2}}= \int_{k'= k -2\pi W}^{k +2\pi W} \left[{\frac{\modu{\tilde v^{\alpha}(k-k')}}{\sqrt{\lambda_{\alpha}}}}\right]^{2} P(k') d k'
\eeq
i.e. it is the unbiased estimator of the true power spectrum smoothed by a strict bandpass filter of width $4\pi W$. Therefore, in a multitaper setting, when combining different tapers with weights, the weights should be chosen such that the eigenspectrum obtained by each of the weighted tapers is as close as possible to this ideal estimate. This forms the basis of the AMTM, where one replaces the scalar weights by frequency dependent weight functions $b_{\alpha}(k)$  which minimize the mean squared error,
\beq
\mathrm{MSE}^\alpha(k)=\av{\modu{\tilde {\cal Y}^{\alpha}(k)-b_\alpha(k) \tilde y^\alpha(k)}^2} .
\eeq
The minimization leads to the following expression for $b_{\alpha}$, 
\beq
\label{bms}
b_\alpha(k)=\frac{\sqrt{\lambda_{\alpha}} P(k)}{\lambda_\alpha P(k)+(1-\lambda_\alpha)\sigma^2}
\eeq
where, $\sigma^2=\int_{-k_N}^{k_N} \frac{dk }{2\pi} P(k)=\mathrm{var}\{x_k\}$, the variance of the map, using Parseval's theorem.\par
With this the minimum mean square error becomes,
\beq
\mathrm{MSE}^\alpha(k)\simeq \frac{P(k)(1-\lambda_\alpha)\sigma^2}{\lambda_\alpha P(k)+(1-\lambda_\alpha)\sigma^2},
\eeq
and the best estimate of the power spectrum has the form,
\beq
\label{amtmSpec}
\hat P^{\mathrm{AMTM}}(k)=\frac{\sum_{\alpha=0}^{M-1} b^{2}_{\alpha}(k)\hat P^{\alpha}(k)}{\sum_{\alpha=0}^{K-1} b^{2}_{\alpha}(k)}
\eeq
where $\hat P^{\alpha}$ is the single taper power spectrum estimate with the $\alpha^{th}$ taper i.e. the eigenspectrum of order $\alpha$. In case of a white noise spectrum $P(k)=\sigma^2$, and therefore $b_\alpha(k)=\sqrt{\lambda_{\alpha}}$, which gives us the formula \eqref{mtmSpec} for the simple MTM. Also, in this case $\mse^\alpha=(1-\lambda_\alpha)\sigma^2$. This shows again that for white noise, for which the first $2NW-1$ tapers have eigenvalues close to unity, the mean squared error is negligible.\par
Note that the estimation of the power spectrum requires the evaluation of the optimal value for the weights $b^\alpha(k)$, which assumes knowledge of the true power spectrum. But the latter is precisely what we are trying to estimate. Therefore, this method has to be  implemented iteratively according to the following steps, 
\begin{enumerate}
\item We use the first one or two tapers (having the least spectral leakage) to form a first estimate of the power spectrum $P(k)$ via \eqref{mtmSpec}. 
\item We use \eqref{bms} to estimate the weights $b_{\alpha}(k)$.
\item With the $b_{\alpha}$'s estimated, we use \eqref{amtmSpec} with $M \lesssim 2NW-1$ tapers to get the second estimate of the power spectrum.
\item Using this new estimate, re-evaluate the weights $b_{\alpha}(k)$ and repeat steps 2-4.
\end{enumerate}
In the following, we will generalize the AMTM to two dimensions.
\subsection{AMTM in two dimensions}
 \begin{figure*}[t]
   \includegraphics[scale=0.6, angle=0]{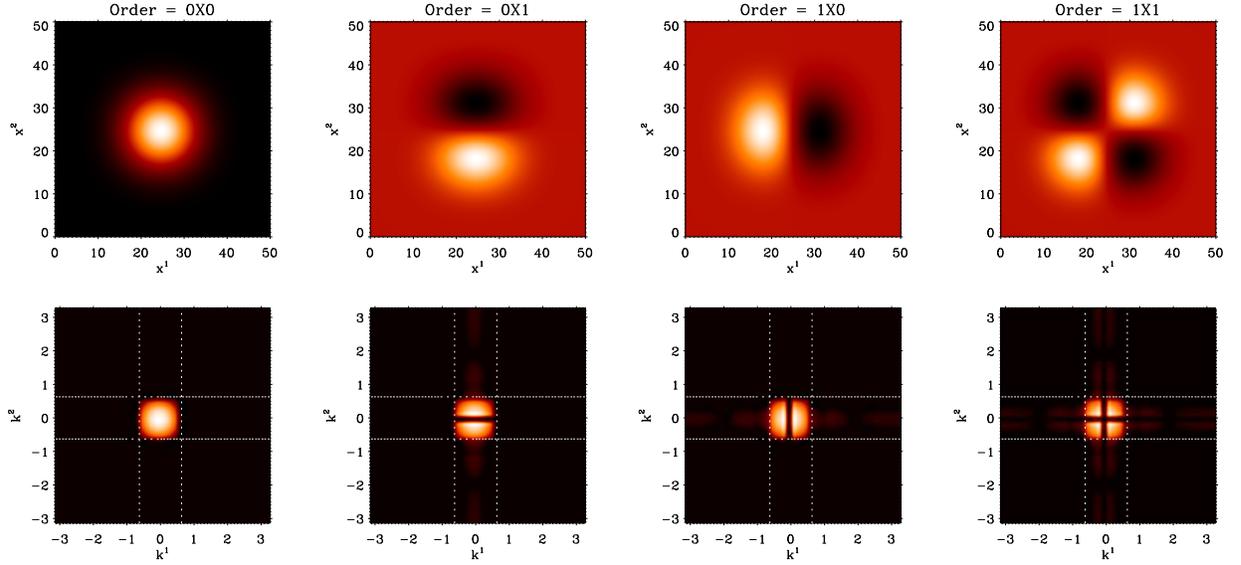}
   \caption{\label{2dtapers} \emph{Top panel}: The first 4 two-dimensional tapers on a N$\times$N grid with $N=50$ and $NW=5$. The order of the one-dimensional tapers corresponding to each taper is indicated on the top. \emph{Bottom panel}: The logarithm of the spectral window functions $\Gamma(k^{1},k^{2})$, corresponding to the tapers on the upper panel.  The color scale on the lower panel are standardized to be $(10^{-30},1)$ times the maximum in each plot. The white dotted lines represent the location of  $\pm 2\pi W$ wavenumbers.} 
 \end{figure*}
So far we have only considered one-dimensional tapers. In the signal processing literature,  generalizations of the multitaper method to higher dimensions have not been widely discussed. Some of the early works include \cite{1988:Barr, 1988:Bronez, 1992:Liu}. Relatively recently, straightforward generalizations of the method to higher dimensional euclidean spaces \citep{1997:Hanssen} and to patches on the surface of the sphere \citep{2005:Wieczorek, 2007:Dahlen, 2004:SDW} have been formulated. In view of the upcoming CMB experiments (that will map the sky on small scales at high resolution), we will be mostly interested in the power spectrum estimation on regularly sampled two-dimensional flat spaces like a projection of a small patch on the sky. As such we will discuss the 2-D extension of the multitaper method as discussed in \cite{1997:Hanssen}.\par
Two-dimensional tapers are constructed from outer products of one-dimensional tapers. We assume a two-dimensional map spanned by the coordinates $(x^{1},x^{2})$ . The data should be sampled regularly, but the sampling intervals (pixel sizes), $\Delta x^{i}$,  and the number of pixels  in each direction, $N_i$, can be different ($i=1,2$).  Let us relabel the one-dimensional taper of the previous section as $\vec v^{\alpha;N}$ where the extra superscript denotes the number of pixels in a given direction. Then a two-dimensional taper of order $(\alpha_1;\alpha_2)$ and size $N_1\times N_2$ can be constructed out of the outer product of two one-dimensional tapers:
\be
\mathbb V^{(\alpha_1\alpha_2;N_1N_2)} = \vec v^{(\alpha_1,N_1)}\left[\vec v^{(\alpha_2,N_2)}\right]^T ,
\ee
where we have treated $\vec v$ as a column vector.
The spectral concentration eigenvalue of a two-dimensional taper, is easily shown to be the product of the eigenvalues of the one-dimensional tapers out of which the former is constructed: 
\be
\lambda_{(\alpha_1\alpha_2;N_1N_2)}=\lambda_{\alpha_1,N_1}\lambda_{\alpha_2,N_2}.
\ee
Two-dimensional tapers constructed in this way inherit most of the properties of one-dimensional tapers such as orthogonality on the sample plane, and optimal spectral concentration and orthonormality in the frequency plane. The two-dimensional power spectrum estimator corresponding to these tapers is defined, similar to the one-dimensional case, as a weighted sum of approximately independent tapered power spectra 
\be
\label{AMTM2D}
P^\mathrm{AMTM}(k_1, k_2)=\frac{\sum_{\alpha_1,\alpha_2} b^2_{(\alpha_1\alpha_2)}(k_1,k_2) \hat{P}^{(\alpha_1\alpha_2)}(k_1, k_2)}{\sum_{\alpha_1, \alpha_2}b^2_{(\alpha_1\alpha_2)}},
\ee
where we have dropped the  $N_{1} N_{2}$ portion of the labels for simplicity. Note that in the above formula the weights $b_{(\alpha_{1}\alpha_{2})}$ depend on the eigenvalues $\lambda_{(\alpha_{1}\alpha_{2})}$  and are given by an equation analogous  to \eqref{bms} and are to be estimated iteratively. The quantity 
$(2\pi)^{2 }\hat{P}^{(\alpha_1\alpha_2)}(k_1, k_2)$ is the eigenspectrum of order $(\alpha_1\alpha_2)$  and is given by 
\be
\hat{\mathbb P}^{(\alpha_1\alpha_2)}= \modu{\mathrm{FT}\left[\mathbb V^{(\alpha_1\alpha_2;N_1N_2)}\mathbb  M^{T}\right]}^2,
\ee
where $\mathbb M =\left\{M(x_{i},x_{j})\right\}$ is the two-dimensional map, the  power spectrum of which is being estimated. \par 
We will introduce a bit of notation at this point. We will designate the two  parameters that control the multitaper method as: 
\begin{itemize}
\item $N_{tap}$: The number of tapers used. Its value will be written in the form $M^{2}$, where $M$ denote the number of one-dimensional tapers. For example if $2$ one-dimensional tapers are used to create $4$ two-dimensional tapers via outer products, then we will quote $N_{tap} = 2^{2}$.
\item $N_{res}$: We will use this as the shorthand for the resolution parameter $NW$. For  example, $N_{res}=3$ will mean that the half-bandpass chosen for generating the tapers is three times the fundamental frequency. 
\end{itemize}

\subsection{Statistical Properties \label{statProp}}
Statistical distribution of the power spectra of a Gaussian random field realized on a map with periodic boundary conditions, such as a full sky CMB field, can be described in terms of the simple analytic distributions. This property stems from the fact that each Fourier-mode (or spherical harmonic component) of such a map is a statically independent quantity. On a finite patch of the sky or for any map with a non-periodic boundary condition, these modes get entangled due to the convolution with the window, and are no longer amenable to such simple descriptions. As we will discuss in this 
subsection, the multitaper method approximately restores many of these nice properties of random fields on a finite patch, and makes the statistical properties  of the spectral estimators describable via simple and intuitive analytic expressions.\par
Most of the statistical properties of the different spectral estimators that  we have discussed so far stem from the basic result that for most stationary processes with a power spectrum $P(k)$ that is continuous over the interval $[-k_{N},~k_{N}]$, the simple FFT power spectrum (periodogram) $P^{PM}(k)$ is distributed as,
\beq
\label{periodogramPDF}
\hat P^{PM}(k)  \eqd 
\begin{cases}
P(k) \chi^{2}_{2}/2, & \text{for $0<k<k_{N}$}; \\
P(k) \chi^{2}_{1}, & \text{for $k=0$ or $k=k_{N}$},
\end{cases} 
\eeq
asymptotically as $N\rightarrow \infty$. Here $\eqd$ means ``equal in distribution'', which means that the statement ``$X \eqd a \chi^{2}_{\nu}$'' is equivalent to saying that the random variable $X$ has the same distribution as a chi-square random variable with $\nu$ degrees-of-freedom (dof) that has been multiplied by a constant $a$. For a Gaussian white noise process the above result is exact for any $N$. Also, for the asymptotic case, the power spectra for two different frequencies $k$ and $k'$ are uncorrelated. Although these results are true for asymptotically large $N$, in a finite $N$ case, they approximately hold for the $N/2+1$ independent Fourier frequencies $k_{j} = (2\pi j)/(N\Delta x)$, if $N$ is large enough. \par
Using the above result, it is possible to predict the approximate distribution of an AMTM spectrum estimator \eqref{AMTM2D}. Using an equivalent degrees-of-freedom argument (see appendix~\ref{appendixZ} for details), it can be shown that for $N\rightarrow \infty$,
\beq
P^{\mathrm{AMTM}}(\bk) \eqd  \frac{\av{P^{\mathrm{AMTM}}(\bk)}}{\nu(\bk)} \chi^{2}_{\nu}
\eeq
where,
\beq
\label{amtmNu}
\nu(\bk) = \frac{2\left(\sum_{\alpha_1, \alpha_2}b^2_{(\alpha_1\alpha_2)}(\bk)\right)^{2}}{\sum_{\alpha_1,\alpha_2} b^4_{(\alpha_1\alpha_2)}(\bk)}.
\eeq
Therefore, for finite $N$, it is reasonable to expect that  the AMTM power spectrum at each pixel is approximately distributed as $\av{P^{\mathrm{AMTM}}} \chi^{2}_{\nu}/\nu$ with $\nu$ given by the above equation. Note that in the case of MTM, $b^{2}_{\alpha_{1}\alpha_{2}} = \lambda_{\alpha_{1}\alpha_{2}}$. If we are only using tapers with eigenvalues close to unity, then $\nu(\bk) \simeq 2M$, $M$ being the total number of tapers used.
\par
Now we turn to the approximate form of the distribution for the power spectrum after it is binned in annular rings in $\bk$ space, which we denote by $P^{B}(k_{b})$,
\beq
\label{binnedPower}
P^{B}(k_{b}) = \frac{1}{N_{b}}\sum_{i,j \in b} P^{\mathrm{AMTM}}(k_{i},k_{j})
\eeq
 where the sum is over all pixels that fall inside bin $b$ and $N_{b}$ is the number of observations in that bin. Using a similar argument as before, it can be shown (see appendix~\ref{appendixZ}) that,
\beq
 P^{B}(k_{b}) \eqd \frac{\av{P^{B}(k_{b})}}{\nu_{b}} \chi^{2}_{\nu_{b}}
\eeq
where the degree-of-freedom, $\nu_{b}$ is given by,
\beq
\label{binnedPDFNu}
 \frac 1{\nu_{b}} \simeq \frac{2 N_{res}^2} {N_{b}^{2}} \sum_{i,j \in b} \frac{1}{\nu(k_{i},k_{j})},
\eeq 
with $\nu(\bk)$ given by \eqref{amtmNu}.
 If the degree-of-freedom variable is also slowly varying, then this implies $\nu_{b} = N_{b}/(2 N_{res}^2) ~ \nu(\modu{\bk} \simeq k_{b})$, which is the expected result for the sum of  $N_{b}/(2 N_{res}^2)$  independent identically distributed $\chi^{2}_{\nu}$ variables. Note that  the appearance of the $\nres^{2}$ factor essentially arises from the fact that AMTM significantly correlates $\nres^{2}$ nearby pixels, and therefore only $N_{b}/(2 \nres^{2})$ ``super-pixels'' are approximately statistically independent. \par
In case of the periodogram, $\nu = 2$ and $N_{res}=1$, so that $\nu_b \simeq N_b$. In case of MTM with $M$ tapers with good leakage properties, $\nu \simeq 2 M$ and therefore, $\nu_b \simeq N_b M/\nres^{2} $.
\section{Application to the ideal CMB map \label{application}}
\begin{figure}[htbp]
\begin{center}
\includegraphics[scale=0.4]{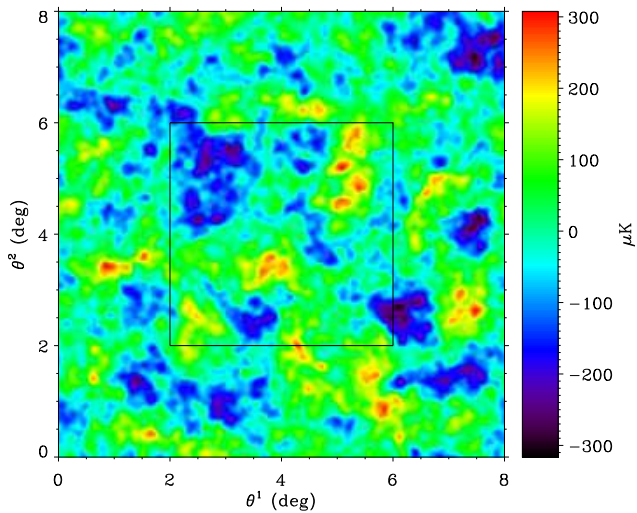}
\caption{One realization of the CMB map on a $192\times 192$ pixel grid. The physical size is 8 degrees on a side. Estimation of the power spectrum is done on the 4 degree $\times$ 4 degree subarea indicated by the rectangle. The color scale represents temperature fluctuations in micro-Kelvin.}
\label{cmbMap}
\end{center}
\end{figure}

In this section, we will illustrate the multitaper method by applying it to flat-sky cosmic microwave background  (CMB) maps.   We shall assume the maps to be pure CMB without noise or masks and with uniform weight. We also assume a square map with square pixels for simplicity, although all the results shown hold for non-square maps and/or non-square pixels. \par

\begin{figure*}[htbp]
\begin{center}
\includegraphics[scale=0.6]{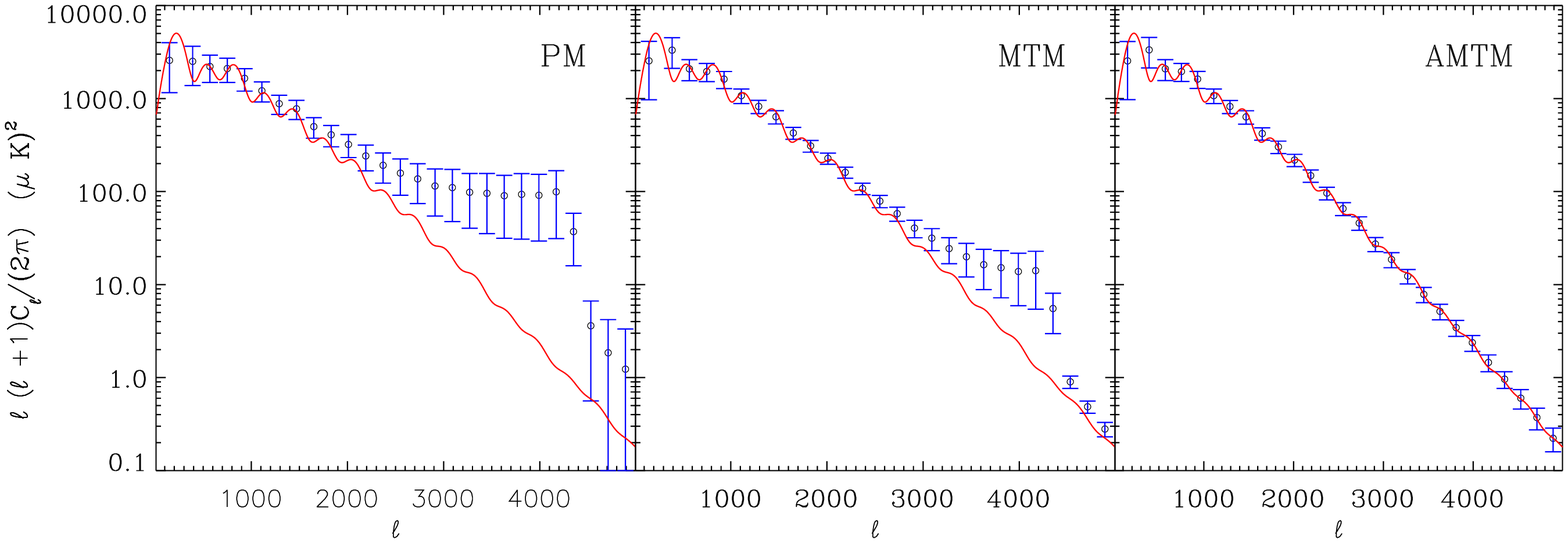}
\caption{Comparison of the simple periodogram method (labeled PM), the eigenvalue weighted multitaper (labeled MTM) and the adaptive multitaper (labeled AMTM) methods for estimating the power spectrum of a CMB map. In each plot, the  continuous line represents the theory power spectrum used as an input for the Monte Carlo simulations. The open circles represent the mean values in each $\ell$ bin, averaged over $5000$ realizations, while the vertical lines represent the $2\sigma$ spread. The bin width for this figure was $\Delta \ell = 180$. For the MTM and AMTM methods $N_{tap}=3^{2}$ tapers with $N_{res} = 2$ were used (see text for details). Note that the power spectra for the multitaper methods appear smoothed because they are convolved with the window function of an effective taper. Standard decorrelation techniques, like the MASTER algorithm \citep{2002:Hivon} can be employed to de-bias and deconvolve all the above power spectra, but in the first two cases, where the mode-coupled power spectra are biased, decorrelation leads to bigger uncertainties in the deconvolved power spectra (see \S~\ref{master}). }
\label{monteCarloPS}
\end{center}
\end{figure*}
We adopt a deliberate change of notation at this point to make the following sections more compatible with existing CMB literature. Since we will dealing with angular co-ordinates on the sky, we call the real space variable $\btheta$ rather than $\bx$. We also call the dual Fourier space, the $\bell$ space, instead of the $\bk$ space. The latter is motivated by the fact that the full sky CMB power spectra are expressed in terms of multipole moments, which are denoted by $\ell$, and the $\bk$ vector is the correct generalization of the $\ell$ modes in the flat-sky case. We will also denote the power spectrum $P(k)$ by $\cl$.\par 
We generate the CMB maps as Gaussian random realizations from a theoretical power spectrum $\cl$. In order to simulate the effect of non-periodic boundary conditions we first generate larger maps from which the desired region is extracted. In the present example, we generated $5000$ realizations of $192\times 192$ pixel CMB map having a physical size of $8$ degrees on a side.  This implies that the sampling intervals (i.e. pixel scales) along either axes is $\delta\theta = 2.5^{\prime}$. We extract a $96\times 96$ pixel 4 degree square sub-map from the center of each such realization and perform the power spectral estimation on this sub-area (see Fig.~\ref{cmbMap}). Given the physical size and  the number of pixels, one finds the Nyquist and fundamental values of $\ell$ to be $\ell_{Nyq} = \pi/(\Delta\theta) = 4320 $ and $\ell_{fund} = (2\pi)/4^{\circ} = 90$.  Therefore, if we choose a resolution parameter of $N_{res} =2$ the half-bandwidth outside which spectral leakage is minimized is $W_\ell = 180$. \par
 We perform the straight FFT or periodogram (labeled PM), MTM and AMTM power spectrum estimation on $5000$ random realizations of the CMB map. The results are shown in Fig~\ref{monteCarloPS}. These plots show the mean power spectrum  binned uniformly in $\ell$ in bins of $\Delta \ell = 180$. The error bars correspond to the $2\sigma$ spread in the distribution of their values. Two features are  immediately apparent from this figure. First, it shows that the PM has significant spectral leakage and produces a power spectrum which is highly biased. Although the eigenvalue weighted MTM is better in this respect, it still suffers from significant bias at  high $\ell$'s because the higher order tapers cannot guard efficiently against spectral leakage. The AMTM seems to do very well in minimizing spectral leakage and producing an approximately unbiased estimate of the input power spectrum. Second, the $2\sigma$ spread in the uncertainty of the binned value in the PM case is much higher than the corresponding spread for the AMTM. This is one of the main reasons for performing tapered power spectrum estimation with multiple tapers, as has been stressed before. As discussed in \S~\ref{master} this property becomes extremely important when the tapered power spectra are decorrelated via a MASTER-like \citep{2002:Hivon} algorithm. The nearly unbiased nature of the AMTM power spectrum translates to errorbars in the deconvolved spectrum which are several factors smaller than those obtained from a periodogram in the large $\ell$ regime. \par
Now, we will compare  the distributions of the various power spectrum estimators with the theoretical expectations of section~\ref{statProp}. For this purpose, we choose three multipoles $\ell\sim 1000, 2000, 3000$ at which we study both the single pixel and the binned probability distributions. The reason for choosing these three numbers is that the first multipole is an example where all three methods are approximately unbiased, the second multipole is a case where the PM is biased but the MTM and AMTM are almost unbiased, while the third multipole represents a regime where only the AMTM is close to an unbiased estimate. \par
\begin{figure*}
\includegraphics[scale=0.6]{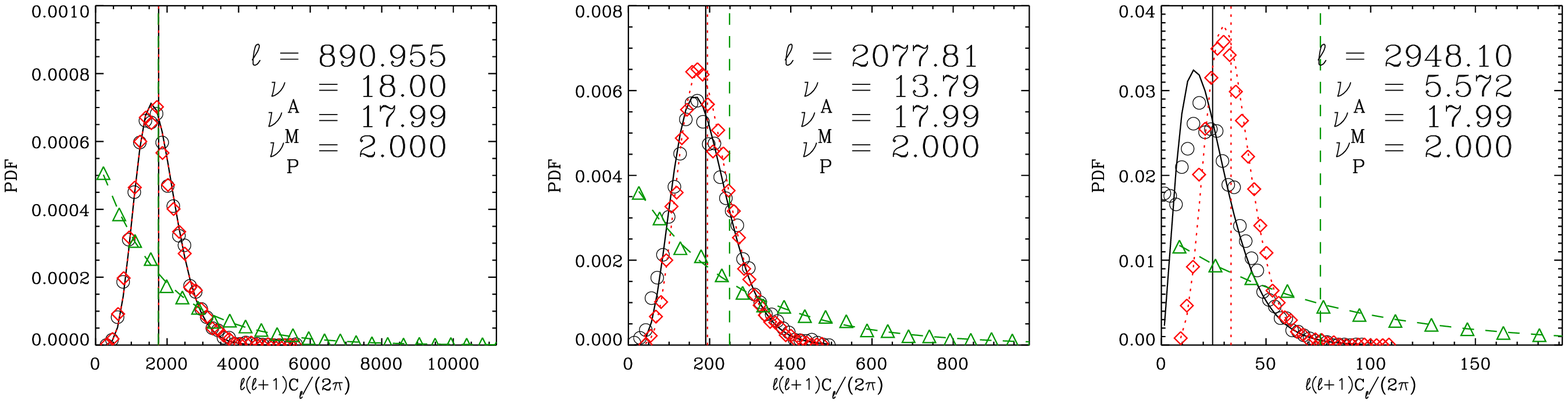}

\includegraphics[scale=0.6]{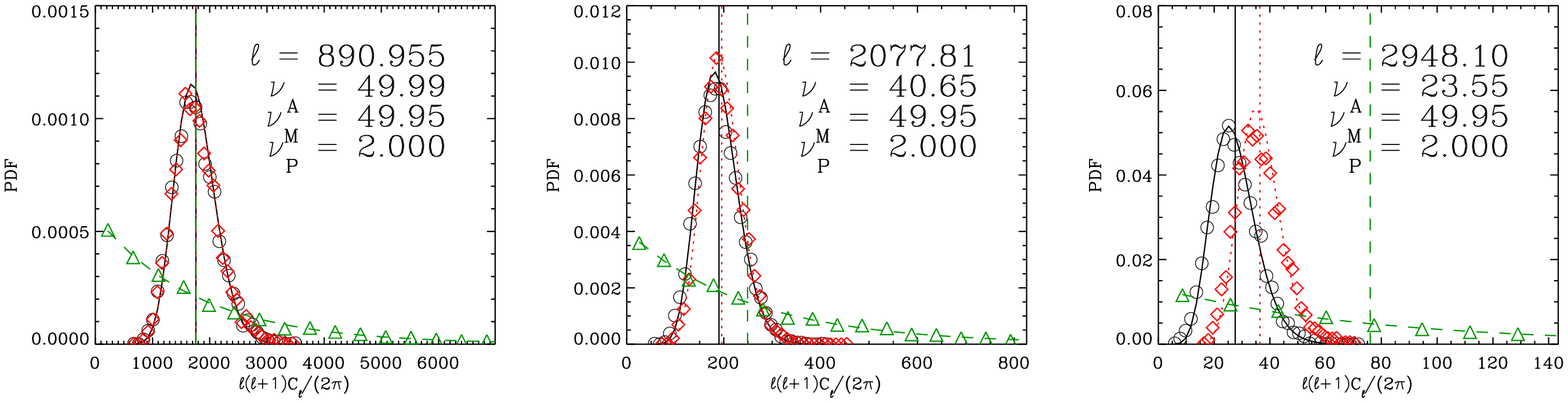}
\caption{\label{singlePixelPDF} Comparison of the probability distributions of the estimated power spectrum at three points (pixels) in the two-dimensional Fourier ($\bell$) space. From left to right, these points are $(\ell^{1},\ell^{2}) = (630,630)$, $(1980,630)$ and $(2880,630)$, for which the modulus $\ell$ has been indicated in each figure. \emph{Upper Panel:} Power spectrum estimation with $N_{res} = 2.0$ and $N_{tap}= 3^{2}$. The open circles (black), the diamonds (red) and the triangles (green) indicate the probability distribution of the quantity $\ell(\ell+1)\cl/(2\pi)$ as estimated via the AMTM, MTM and PM methods from $5000$ Monte Carlo simulations, respectively. The respective approximate theoretical forms as discussed in \S~\ref{statProp} are over-plotted as the continuous curve (black), the dotted curve (red) and the dashed curve (green) for each of the methods. The mean degree of freedom of the chi-square for each method is also indicated as $\nu_{A}$ for AMTM, $\nu_{M}$ for MTM and $\nu_{P}$ for the periodogram, PM. Each curve is also accompanied by a vertical line of the same style (and color)  representing the mean value obtained from the Monte Carlo simulations. In each figure, the continuous (black) vertical line corresponding to the mean of the AMTM method, is also the value closest to the true power spectrum. It is actually the unbiased value of the pseudo power spectrum (see \S~\ref{master}). {\emph {Lower Panel:}} Same as above but with $N_{res}=3.0$ and $Ntap=5^{2}$.}
\end{figure*}

For the single pixel case, we choose three pixels $(\ell^{i},\ell^{j})$ on the $\bell$-space  map, such that values of $\modu{\bell}$ are close to the three multipoles as described above. We store the values of $\cl$ realized in the $5000$ Monte Carlo simulations in these pixels, and draw up the probability distribution of the quantity $\ell(\ell+1)\cl/(2\pi)$ from these values, for each of the three methods. The results are shown in Fig~\ref{singlePixelPDF}. In the top panel, we show the results of the same experiment that was used to generate Fig.~\ref{monteCarloPS}, i.e. with $N_{res}=2.0$ and $N_{tap}=3^{2}$. For the leftmost plot, all three methods of power spectrum estimation are approximately unbiased. The MTM and AMTM methods are almost identical here, because the spectrum does not have a large local slope at this multipole so that even in the adaptive method, the higher order tapers do not need to be down-weighted to reduce spectral leakage. Therefore, the equivalent degrees-of-freedom come out to be the same and $\simeq 2N_{tap} = 2\times 3^{2} = 18,$ as expected for a white spectrum from \eqref{amtmNu}. For the middle plot, the multipole $\ell$ is such that the local spectrum is moderately colored and spectral leakage for a periodogram becomes apparent in the biased estimate of the mean power spectrum it generates, as indicated by the dashed vertical line. The AMTM remains approximately unbiased while the MTM is only slightly biased, but the equivalent degrees-of-freedom for AMTM is now lower than that of the MTM method. This is expected because to reduce spectral leakage the weights associated with the higher order tapers have been reduced in the AMTM spectrum, thereby lowering the degrees-of-freedom. This trend continues to higher multipoles, and as shown in the rightmost plot of in the figure, the  spectral leakage is kept at a minimum only in the AMTM power spectrum by heavily down-weighting the higher order tapers, while both MTM and PM become highly biased. An important point to note here is that not only does AMTM minimize the spectral leakage (and hence bias) but it also takes advantage of multiple tapers by reducing the error in the estimate. This is most easily seen by comparing the width of the distributions of AMTM spectra against that of the PM spectra. The lower panel of Fig.~\ref{singlePixelPDF} essentially shows the same features but for $N_{res} = 3.0$ and $N_{tap}=5^{2}$. A comparison of the upper and lower panels also illustrate how by using more tapers the uncertainty in the power spectrum can be reduced. We would like to remind the reader once again here that the number of usable tapers depend on the spectral resolution chosen - poorer spectral resolution (in this example $N_{res} = 3$ vs. $2$) allows the use of more tapers ($N_{tap} = 5$ vs. 3) and consequently an estimate with lower error.\par
\begin{figure*}
\includegraphics[scale=0.6]{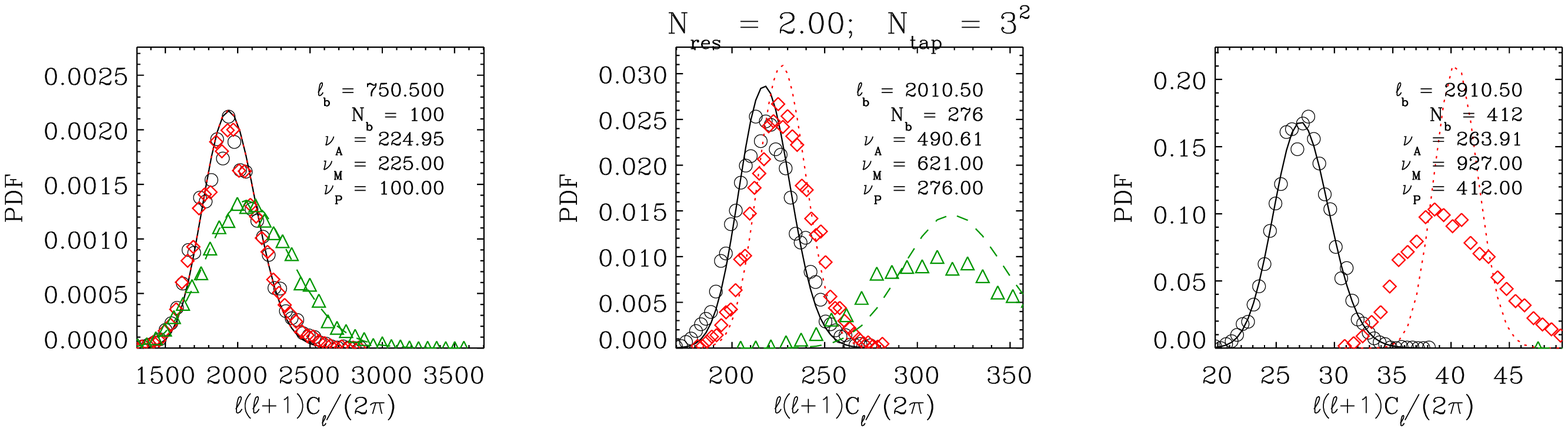}

\includegraphics[scale=0.6]{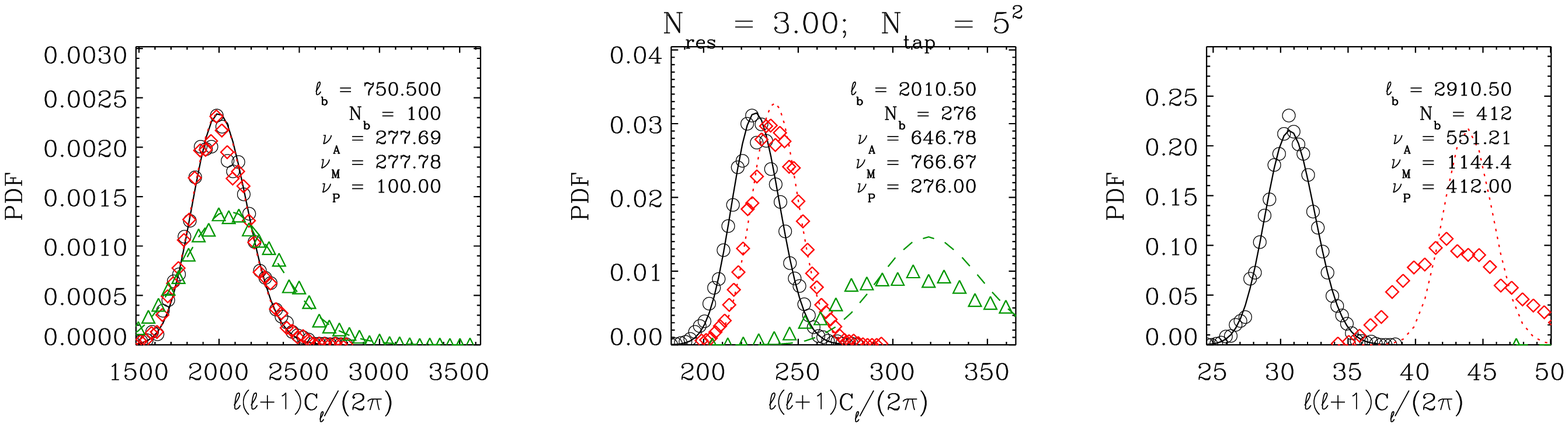}

\includegraphics[scale=0.6]{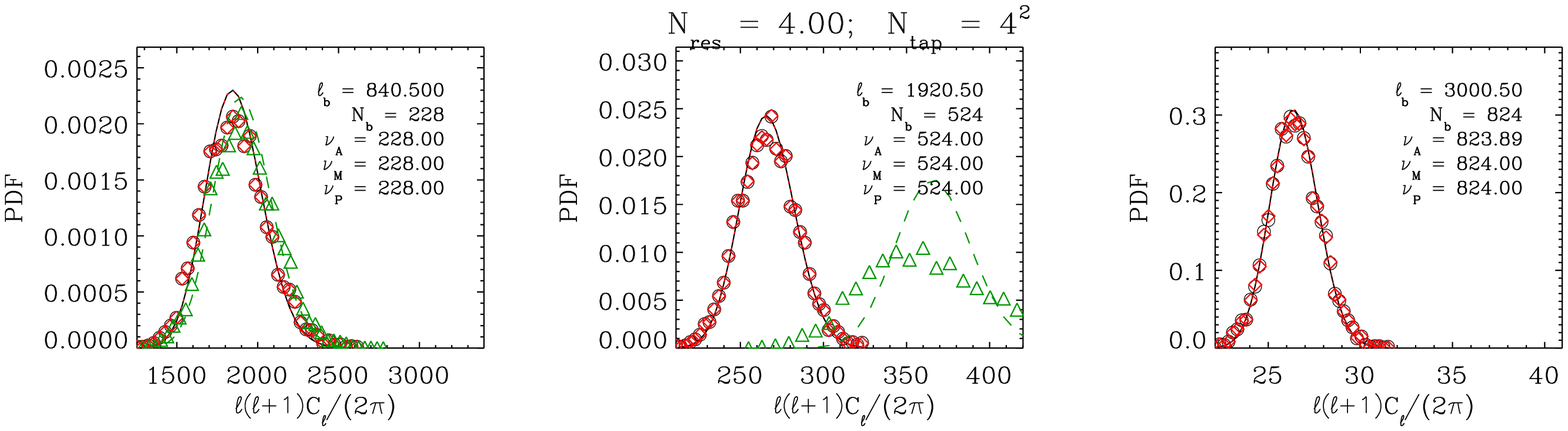}
\caption{\label{binnedPDF} Comparison of the probability distributions of the estimated power spectrum in bins. Each panel is for a different combination of the $N_{res}$ and $N_{tap}$ as indicated on the top of the middle figure. The open circles (black), the diamonds (red) and the triangles (green) indicate the probability distribution of the quantity $\ell(\ell+1)\cl/(2\pi)$ in each bin as estimated via the AMTM, MTM and PM methods from $5000$ Monte Carlo simulations, respectively. The respective approximate theoretical forms as discussed in \S~\ref{statProp} (see eq.~\eqref{binnedPDFNu}) are over-plotted as the continuous curve (black), the dotted curve (red) and the dashed curve (green) for each of the methods. The mean degree of freedom of the chi-square for each method is also indicated as $\nu_{A}$ for AMTM, $\nu_{M}$ for MTM and $\nu_{P}$ for the periodogram, PM. The number of pixels in each bin is also indicated as $N_b$. Note the the periodogram is absent in each of the rightmost plots, as it is highly biased and lies outside the range plotted.}
\end{figure*}

Now, we turn to the distribution of the bandpowers $C_{b}$, i.e. the mean power spectrum in annular bands in the $\bell$ space.  This is illustrated in Fig.~\ref{binnedPDF}. The top two panels correspond to the same multitaper parameters as in Fig.~\ref{singlePixelPDF}, binned uniformly in $\bell$ space with bins of width $\Delta \ell = 180$, while in the third panel we show the case for another experiment with $\nres = 4.0$ and $\ntap = 4$ binned at $\Delta \ell = 360$. The curves plotted over the points are the approximate theoretical distributions expected from \eqref{binnedPDFNu}. We note that the theoretical curves are always a good fit for the AMTM method. This is mainly because of the fact that the AMTM is approximately unbiased at all multipoles, while the MTM or PM become biased except at the lowest multipoles. Since near-unbiasedness was an assumption adopted in deriving the binned distributions, the latter two methods suffer from mismatch with theory wherever they are highly biased. For the  $\nres = 4.0$ case, we chose a bin width of $360$ so as to make it large enough to avoid splitting a super-pixel between bins. Also, note that the distribution of the binned power spectra are close to Gaussian, an expected result, as binning essentially involves combining many variables with almost identical distributions.\par    
\begin{figure*}
\includegraphics[scale=0.3]{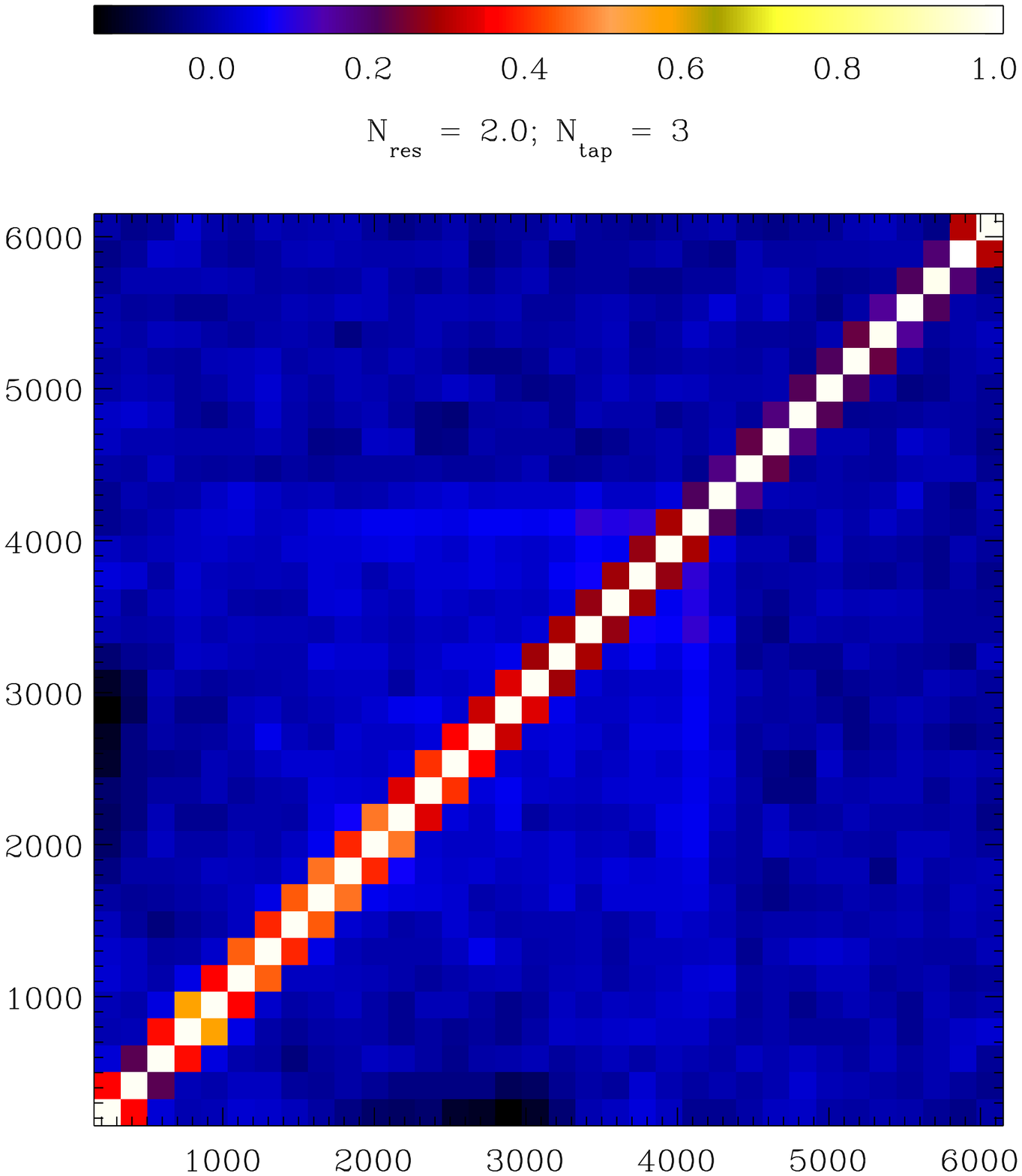}\includegraphics[scale=0.3]{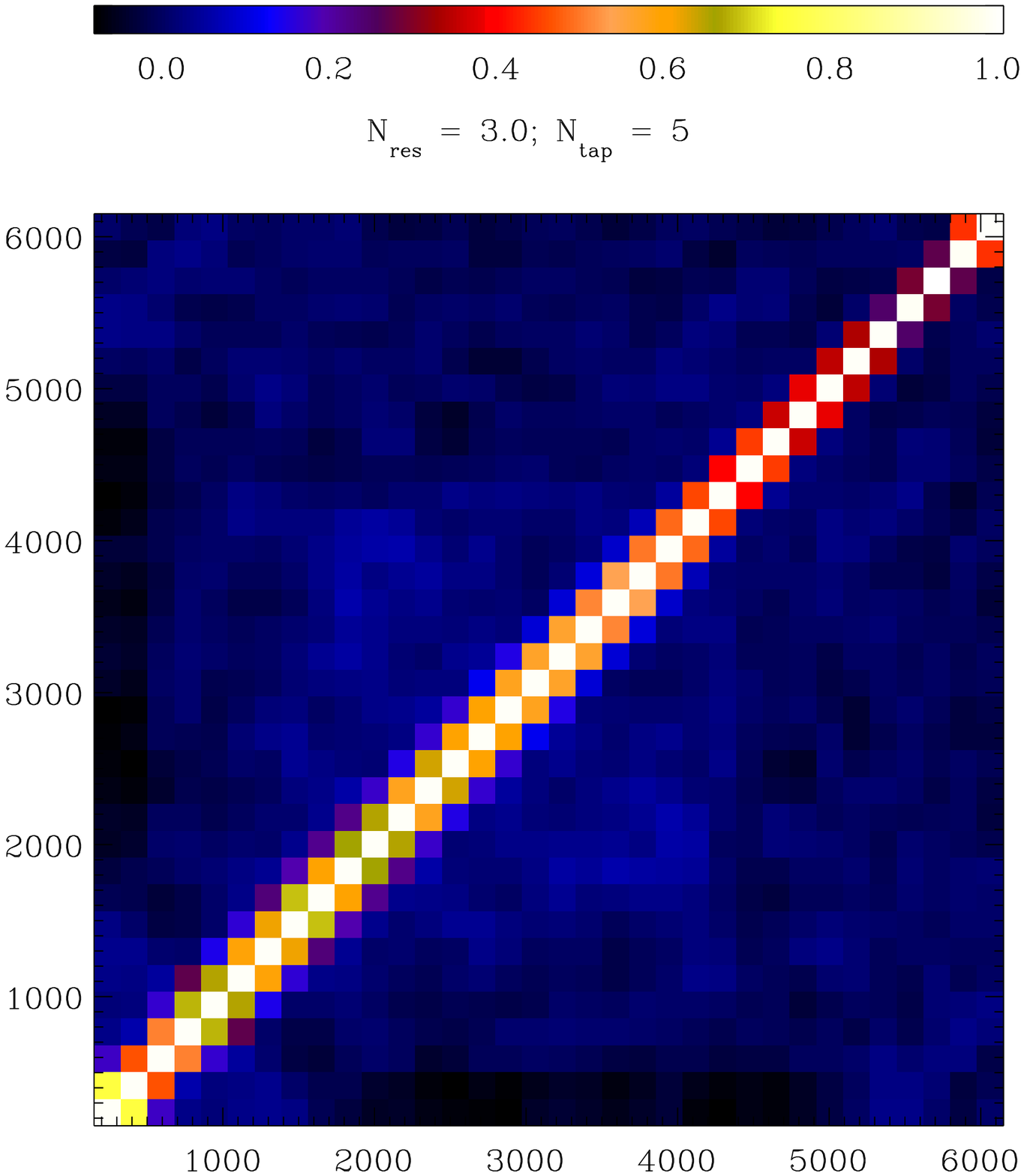}\includegraphics[scale=0.3]{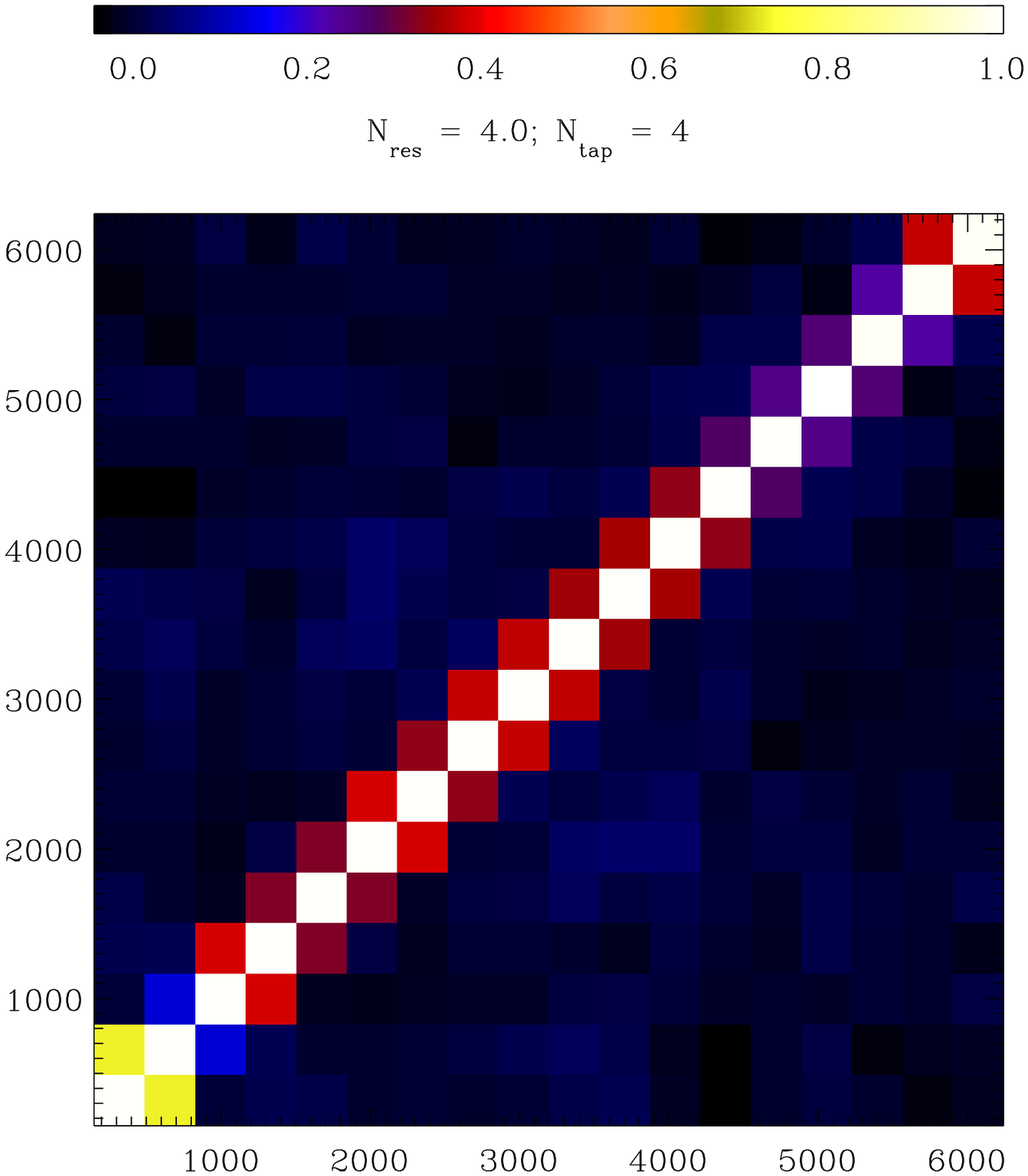}
\caption{\label{covarianceMatrixFig} Covariance matrix of the bandpowers estimated via AMTM for three different parameter settings; from left to right these are: $\nres = 2.0,~~ \ntap =3$; $\nres = 3.0, ~~ \ntap =5$ and $\nres=4.0,~~ \ntap=4$. Each square in the image represents a bin in $\ell$. For the $\nres=2.0$ and $\nres=3.0$ cases, we chose the bin-widths to be twice the fundamental resolution element, i.e. $\Delta \ell = 2 \ell_{fund}= 180$, while for the $\nres=4.0$ case it was taken to be $4 \ell_{fund}= 360$. There is appreciable covariance only  between bins inside the resolution $\Delta \ell_{W} = 2~\nres~ \ell_{fund}$ set by the taper, and the covariance drops drastically beyond that frequency. }
\end{figure*}

Next, we investigate the covariance between different bins, in order to study how the bandpowers are correlated with each other. We define the scaled covariance matrix,
\beq
{\mathcal C}_{a b} =\frac{C_{a b}}{\sqrt{C_{aa} C_{bb}}} 
\eeq
where,
\beq
C_{ab} = \av{(C_a-\av{C_a})^2(C_b-\av{C_b})^2}.
\eeq
We use our Monte Carlo simulations to estimate the quantity above. The results are displayed in Fig.~\ref{covarianceMatrixFig}. These figures illustrate an extremely desirable feature of the AMTM estimator i.e. the bandpowers are essentially uncorrelated beyond the spectral resolution $\Delta \ell_W= 2~\nres~\ell_{fund}$ set by the taper parameters. This implies that if we set the bin widths to be same as $\Delta \ell_W$ then adjacent bins will be uncorrelated. If our bins are smaller, they will be correlated through a mode coupling matrix which is fairly diagonal, and hence can be easily inverted to decouple them, an issue we will briefly touch upon in \S~\ref{master}.\par
In the following section, we turn to the more practical issue of dealing with CMB maps with noise and point source masks in them.

\section{Prewhitening for CMB Maps with mask and noise \label{prewhiten}}
The maps of the cosmic microwave background produced by any experiment will invariably contain instrumental noise and regions, like bright point sources, that are usually masked out before estimating the power spectrum. The raw sky map also contains other astrophysical contaminants which we will neglect for the purpose of this discussion. \par
Application of point source mask to a map is essentially replacing the pixel values in an area containing each point source with zeros. Masking is therefore equivalent to multiplying the map with a function which is unity everywhere except inside discs of varying sizes, where it is zero. This can also be thought of as successively multiplying the map with a mask for each point source. Taking  power spectrum of the final masked map is therefore equivalent to successively convolving the true power spectrum with power spectra of a series of such single-point-source masks. As the size of the discs will in general vary for each such function, the true power spectrum will be convolved with functions that have power over various ranges of multipoles. Although the multiplication with tapers will guard against aliasing of power due to sharp edge of the map, they will be ineffective against the mixing of power due to a point source mask. We propose here a method for dealing with such issues, which in essence is the following:
\begin{enumerate}
\item Perform local real space convolution of the map with designed kernels so as to make its power spectrum as flat (white) as possible, at least over the range of multipoles which is affected most by aliasing of power due to the point source mask. We refer to this procedure as ``Prewhitening''.
\item Perform an AMTM power spectrum estimation of this prewhitened map in order to minimize any additional aliasing of power due to sharp edges of the map. 
\item As the prewhitening operation was a convolution whose Fourier space form is (preferably analytically) known, divide the power spectrum of the prewhitened map with the Fourier space form of the prewhitening function to recover the power spectrum.
\end{enumerate}
Note that the design of the prewhitening function will be specific to the type of signal being considered. In the following, we will demonstrate the prewhitening of CMB maps, with forms of the prewhitening function specific to the features in the CMB power spectrum. We will first consider a noiseless map, in order to motivate the basic form of the prewhitening operation and then generalize to maps with white noise. 
\subsection{Prewhitening of noiseless CMB maps}
Let us denote the pure CMB  temperature map as $T(\btheta)$. An important feature of the power spectrum of the CMB $\cl$ is that beyond a multipole of $\sim 1000$, it is approximately proportional to $\ell^{-4}$. Such sharp fall in the power makes it extremely prone to aliasing of power across multipoles due to a point source mask. If we perform an operation akin to taking the Laplacian of this map, then we would effectively multiply the power spectrum by $\ell^4$ on all scales, thereby making the processed power spectrum nearly white over the large $\ell$ tail. This would thereby minimize aliasing of power. In the following we propose a method of achieving this, by a combination of two operations which we call ``disc-differencing'' and ``self-injection''. \par
\subsubsection{Disc-differencing}
\begin{figure}
\includegraphics[scale=0.4]{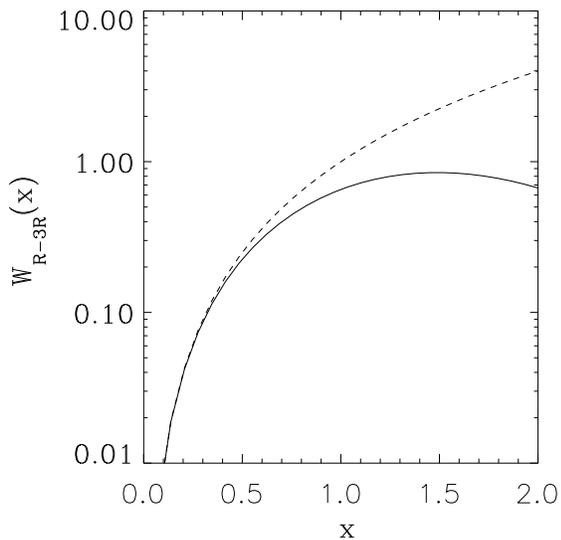}
\caption{``Disc-difference'' function $W_{R-3R}$ discussed in the text (solid line). The dashed line represents the function $x^{4}$.\label{discDiffWindow}}
\end{figure}

If we convolve the map with a circular disc of radius $R$, generating,
\beq
{T_R}(\btheta)=\int d^2\btheta' W_R(\btheta'-\btheta) T(\btheta') 
\eeq
where $W_R$ is the top-hat filter, 
\beq
W_R(\btheta) =
\begin{cases}
\frac{1}{\pi R^2} & \text{if  $\modu{\btheta} \le R$,}
\\
0 & \text{otherwise,}
\end{cases}
\eeq
then in Fourier space, we will have
\beq
T_R(\bell)= W_R(\bell) T(\bell) 
\eeq
where the Fourier space window, $W(\bell)$ is given by,
\beq
W_R(\bell)=2\frac{J_1(\ell R)}{\ell R}.
\eeq
Now let us consider smoothing the map with another top-hat window of radius $3 R$, giving,
\beq
T_{3R}(\btheta)=\int d^2\btheta' W_{3R}(\btheta'-\btheta) T(\btheta').
\eeq
We then take the difference map,
\beq
T_{R-3R}=T_{R}-T_{3R},
\eeq
which in Fourier space reads, 
\beqn
T_{R-3R}(\bell)& = & W_{R-3R}(\ell) T(\ell) \\
& \equiv & 2\l(\frac{J_1(\ell R)}{\ell R}-\frac{J_1(3\ell R)}{3\ell R}\r)T(\ell).
\eeqn
Remembering the asymptotic expansion of $J_1(x)$ for $x<<\sqrt{2}$,
\beq
2\frac{J_1(x)}{x} \simeq 1-\frac{x^2}{8},
\eeq
which shows that for values of  $\ell$ such that   $\ell R <\sqrt{2} $, we have,
\beq
T_{R-3R}(\bell)\simeq (\ell R)^2 T(\ell),
\eeq
which implies that the power spectrum is now,
\beq
(2 \pi)^2\cl^{R-3R} \simeq \av{T_{R-3R}(\bell)^* T_{R-3R}(\bell)}=(2 \pi)^2 \ell^4 R^4 C_\ell,
\eeq
which is the desired form.  The disc-difference window, $W_{R-3R}(\ell R)$ is shown in Fig.~\ref{discDiffWindow}. Beyond $\ell =\sqrt{2}/R$, the function starts falling again. Thus by choosing the radius $R$ judiciously, one can prewhiten a desired range of the power spectrum.  \par
An important aspect of this method is that it is an effective $\mathbb{C}^{-1/2}$ operation on the map ($\mathbb C$ is the covariance matrix), which is manifestly local, and therefore does not suffer from effects due to edges, which is a common problem in case of the full Fourier space operation. Maximum likelihood methods of estimation of power spectra naturally involve the $\mathbb{C}^{-1}$ operation \citep{1997:Tegmark,1999:Oh}.  However, for high resolution experiments, the numerical costs for computing the $\mathbb{C}^{-1}$ matrix can be prohibitively large. Also, such an operation is non-local and mixes modes from masks with the map in a non-trivial way. The method proposed here is approximate but is  simple to implement as it involves only convolutions in real space: as such, its effect can be quantified, propagated or undone very easily. \par
\begin{figure}
\includegraphics[scale=0.45]{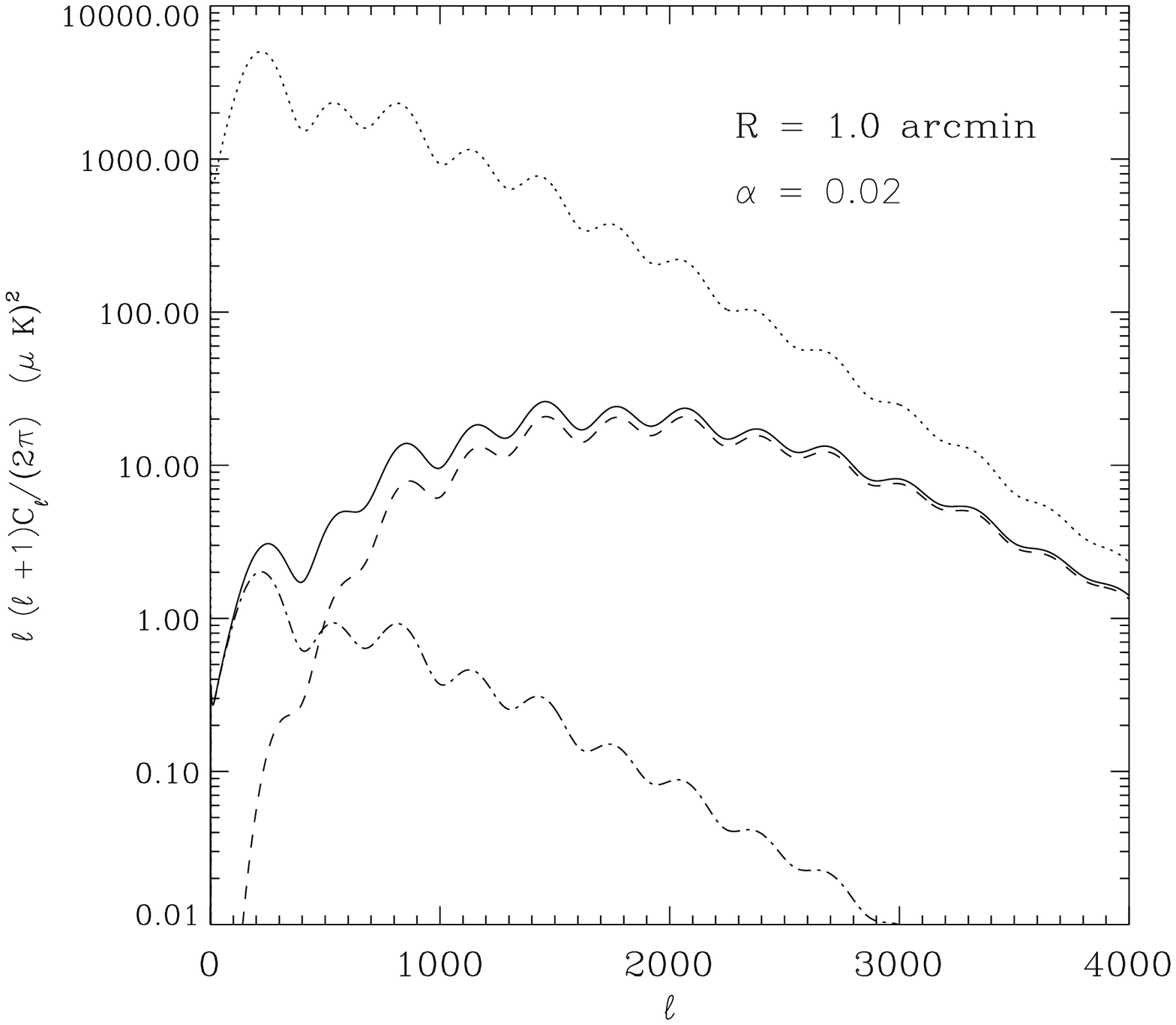}
\caption{\label{prewhiteningTheory} Effect of disc-differencing and self-injection on the power spectrum. The dotted line is the true power spectrum with  a large dynamic range. The disc-differencing operation alone produces the dashed curve which has a much smaller dynamic range, but is steeply rising at low multipoles. The disc radius used is $R= 1^{\prime}$. The dot-dashed curve shows the true power spectrum multiplied by a constant $\alpha^{2}$ where $\alpha=0.02$. If we disc-difference the map followed by self-injection of a fraction $\alpha$ of the map, then the power spectrum of the processed map is the solid curve (given by \eqref{processedSpectrum}) which is conveniently flat over the range of multipoles.}
\end{figure}

Figure~\ref{prewhiteningTheory}  illustrates the effect of the disc-differencing operation on the CMB power spectrum, for a radius $R = 1^{\prime}$. This means that the prewhitening window turns around at $\ell \simeq \sqrt{2}/R \simeq 5000$. The application of the disc-differencing produces the processed power spectrum shown by the dashed curve in the figure, which has much smaller dynamic range than the original one shown by the dotted curve. 
\subsubsection{Self-injection}
One undesirable effect of the application of the disc-differencing function is that it makes the lower multipole part of the CMB power spectrum $(\ell\lesssim 1000)$ a steeply rising function, which may aggravate aliasing of power. A simple way to deal with this problem, is to  add a small fraction of the original map back into the disc-differences map, a process we call ``self-injection''. If a fraction $\alpha\ll 1$ of the map is self-injected after the disc-differencing, the processed power spectrum looks flat at all multipoles, and is conveniently given by a multiplication  of the true power spectrum with an analytic function,
\beq
\label{processedSpectrum}
\cl^{PW} = {(W_{R-3R}(\ell R) + \alpha)^{2}}\cl.
\eeq
This is illustrated for $\alpha=0.02$ in Fig.~\ref{prewhiteningTheory}. \par
Having laid out the basic theory of prewhitening, we will now describe a concrete example of power spectrum estimation of a CMB map with a point source mask to judge the effectiveness of this method in recovering the true power spectrum. To this end we simulate a $8^{\circ} \times 8^{\circ}$ map as a Gaussian random realization from a theory power spectrum. The map has $768$ pixels on a side, making the pixel scale $\Delta\theta = 0.625^{\prime}$. We cut out a  $4^{\circ}\times 4^{\circ}$ section  from this map to impose non-periodic boundary conditions. We then simulate a point source mask for this smaller map, which is unity everywhere except inside randomly positioned discs of various radii, where it is zero.\par
\begin{figure}
\includegraphics[scale=0.38]{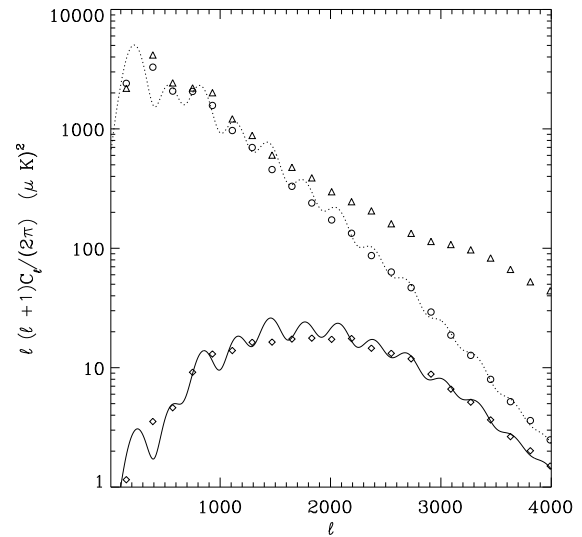}
\caption{\label{prewhitenExampleNoNoise} Prewhitening and AMTM as a remedy to aliasing of power due to point source mask. The dotted curve represents the input power spectrum from which the map is generated. The triangles represent the recovered AMTM power spectrum ($\nres=3.0$, $\ntap=3^{2}$) of the map after a point source mask is applied directly to it. If, on the other hand, the map is first prewhitened (see text) and the mask is subsequently applied, one obtains the diamonds as the AMTM power spectrum. The solid curve is the theoretical prediction for the prewhitened power spectrum. Thereafter, one divides the spectrum by the prewhitening transfer function (see \eqref{processedSpectrum}, obtaining a nearly unbiased estimate, denoted by the open circles. Note that the AMTM power spectra appear smoother than the true spectra as the former is convolved with the window function of the taper. }
\end{figure} 
To motivate the necessity  of prewhitening, we first apply the mask directly to the map and take its AMTM power spectrum with $\nres = 3.0$ and $\ntap = 3^{2}$. The result is shown in Fig~\ref{prewhitenExampleNoNoise}. Note that we expressly use only the first few tapers with least leakage to ensure that mode coupling is minimized due to sharp edges of the map. However, this choice has no bearing against the  aliasing of power due to holes in the map and as expected, we find a lot of power aliased from the low to the high multipoles. Next we perform a disc-difference operation on the original map with $R=1^{\prime}$, followed by a self-injection with $\alpha=0.02$. Then, we apply the mask on this prewhitened map and perform  AMTM power spectrum estimation on it. By comparing with the expected theoretical power spectrum, we find that aliasing is almost non-existent in the power spectrum of the prewhitened and masked map. Then we simply divide this power spectrum by the analytical transfer function for the prewhitening operation \eqref{processedSpectrum}, to obtain a nearly unbiased estimate of the true power spectrum. 
\subsection{Prewhitening of Noisy Maps}
\begin{figure}
\includegraphics[scale=0.38]{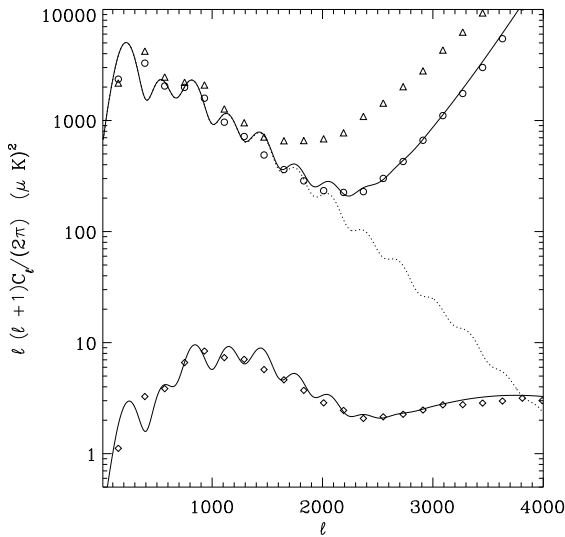}
\caption{\label{prewhitenExampleNoise} Same as in Fig.~\ref{prewhitenExampleNoNoise}, but for a map with white noise. In addition to the standard prewhitening operation, a Gaussian smoothing has been applied to flatten the tail of the prewhitened map.}
\end{figure}

In a real experiment, the map will be convolved with the instrument beam and will contain noise from the instrument as well as other astrophysical signals. To simulate the simplest of such  situations, we convolve the map from the previous step with a Gaussian beam of full-width-at-half-maximum (FWHM) of $5^{\prime}$. Next, we add Gaussian white noise at a level of $3$ $\mu$K per sky pixel (defined as an area of FWHM$^{2}$ on the sky). The power spectrum of the map so generated has the usual initial sharp fall with multipole $\ell$, but then flattens  out beyond the multipole where white noise starts to dominate. One immediate consequence of this is that the disc differencing operation that multiplies the power spectrum by $\ell^{4}$ will make the white noise part rise with $\ell$ instead of remaining flat. One can try to minimize this effect by judiciously choosing the disc radius $R$ so that the disc-differencing window function turns over at the value of $\ell$ where white noise starts to dominate. However, there may be components to the power spectrum other than white noise which may rise faster than the fall of the disc-differencing window. All these cases can be effectively dealt with by introducing another component to the prewhitening operation: namely a convolution with a Gaussian window after the disc-differencing and the self-injection steps. If we convolve the map with a Gaussian of FWHM $\theta_{s}$ then after evaluating the power spectrum we can remove its effect by dividing it by the multipole space form $\exp(\ell(\ell+1)\theta_{s}/(8 \log 2))$. In Fig.~\ref{prewhitenExampleNoise} we illustrate the prewhitening in presence of white noise.  
\section{Mode-mode coupling and deconvolution\label{master}}
\begin{figure*}
\includegraphics[scale=0.38]{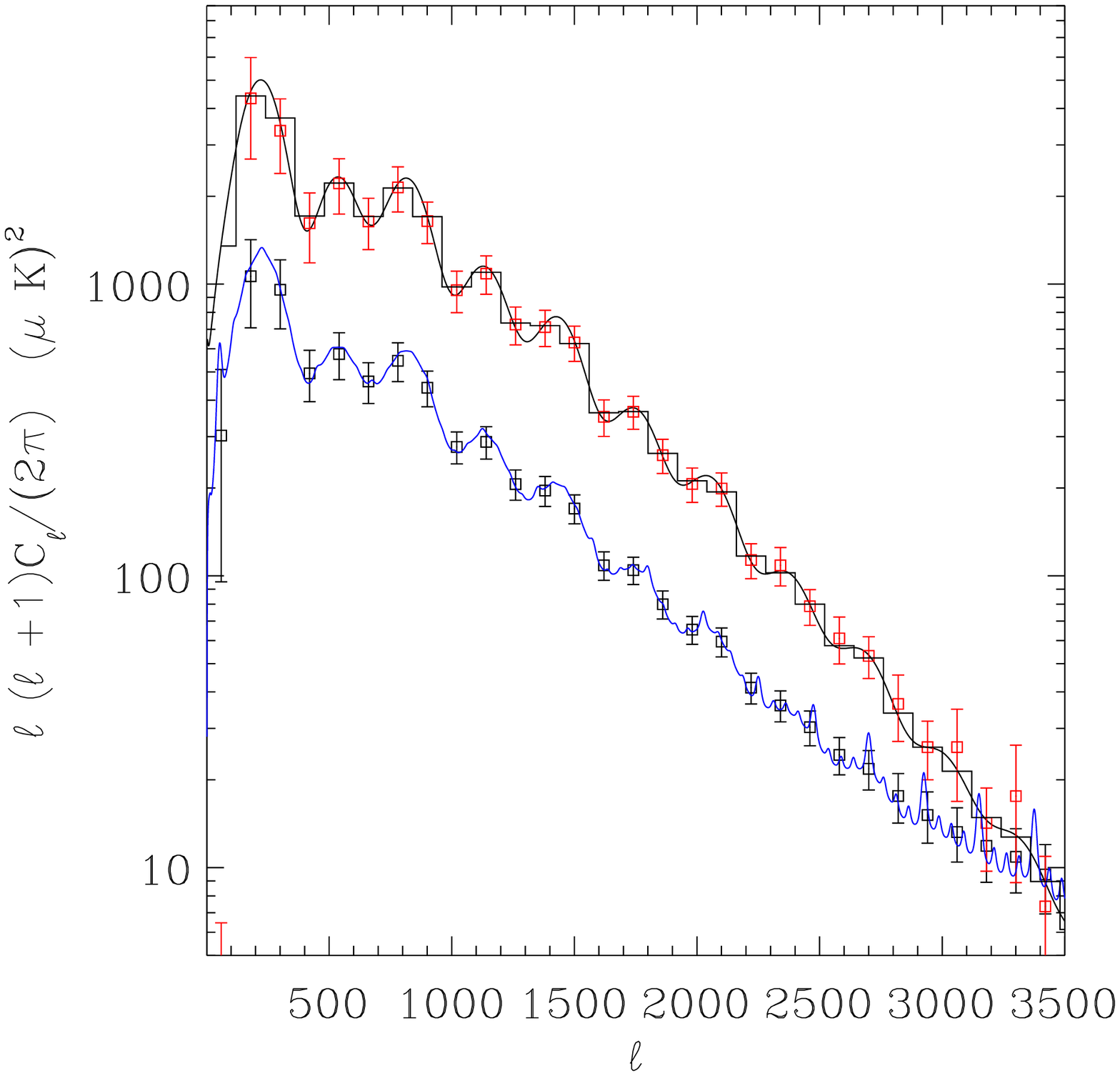}
\includegraphics[scale=0.38]{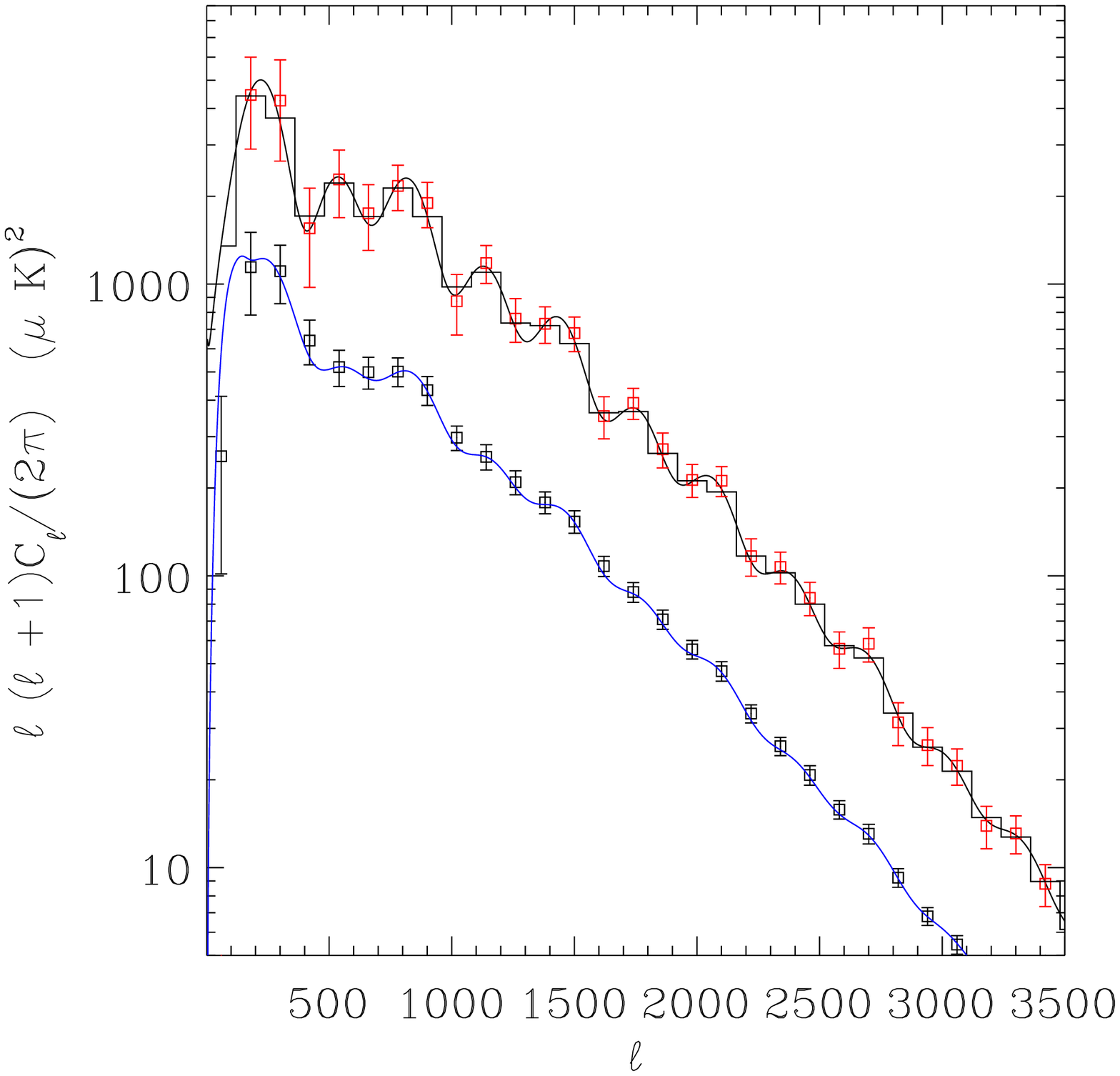}
\caption{\label{slavesAndMasters} Deconvolution of the power spectrum. \emph{Left panel}: Deconvolution of the periodogram with top-hat weighting. The black squares represent the binned power spectrum obtained directly from the map using the periodogram (straight FFT) method. As discussed in the text, this power spectrum is the true power spectrum convolved with the mode-mode coupling matrix due to the top-hat, the theoretical expectation for which is displayed as the blue line. The red points represent the binned power spectrum deconvolved via \eqref{deConvol}. The black line is the input power spectrum. The deconvolved binned power spectrum is to be compared with the binned input power spectrum which is displayed as the black histogram. All points displayed are the mean of $800$ Monte Carlo simulations and the error bars correspond to the $2\sigma$ spread in their values.  {\emph{Right Panel:}} Same as above, but for the AMTM method. The mode-coupled power spectrum and the corresponding theoretical curve in this case have been artificially shifted below the deconvolved power spectra for easy viewing. As discussed in the text, the mode-coupled power spectrum produced by the AMTM method is a nearly unbiased estimate of the true power spectrum, while the  mode-coupled periodogram (left panel) is highly biased at large multipoles. This bias causes the error bars in the deconvolved periodogram to be much larger than those in the deconvolved AMTM power spectrum at large $\ell$, as shown in Fig.~\ref{errorCompare}.}
\end{figure*}
As discussed earlier, an undesirable feature of the Multitaper method is the loss of spectral resolution. For any $\nres>1$, a Multitaper estimator smooths the power spectrum with a frequency-space window which is wider than the fundamental resolution set by the size of the map. For power spectra which are highly structured, this poses the problem of diluting interesting features which may render the power spectrum less useful as a direct probe of the underlying phenomenon. For example, in case of the CMB,  (A)MTM leads to the smoothing of acoustic features. \par
In CMB analyses, the problem of recovering the true power spectrum from one that has been convolved with a window function is a well studied problem. It arises naturally in full-sky CMB experiments because of the application of point source and galactic masks. A detailed account of the procedure to deal with mode-mode coupling in the spherical harmonic space can be found in \cite{2002:Hivon}. In what follows, we will discuss the mode-mode coupling in flat space and develop the method of  deconvolving the effective tapering window function from the power spectrum.\par 
Consider a homogeneous, isotropic and zero-mean Gaussian random process realized on a map, $T(\bx)$, with a power spectrum $P(k)$,
\beq
\av{T(\bk)^{*}T(\bk')} = (2\pi)^{2}\delta(\bk-\bk') P(k).
\eeq
If the map is multiplied by a window $W(\bx)$, then the Fourier transform of the resulting map is given by the convolution,
\beq
T^{W}(\bk) = \int \frac{d^{2}\bk'}{(2\pi)^{2}} T(\bk') W(\bk-\bk').
\eeq
In the ensemble average sense, we therefore have,
\beqn
\label{pseudocl}
\av{\hat P_{W}(\bk) }&\equiv & \frac{1}{(2\pi)^{2}} \av{ T^{W*}(\bk)T^{W}(\bk)}\\
&=&\int\frac{ d^{2} {\bk'}}{(2 \pi)^{2}} \frac{{\modu{W(\bk-\bk')}}^{2}}{(2\pi)^{2}} P(k'),
\eeqn
which is often referred to as the pseudo power spectrum in CMB literature. All the multitapered power spectra discussed so far are pseudo-power spectra in this sense, and comparing with \eqref{AMTM2D}, it is easily seen that for AMTM, the power spectrum of effective window that couples to the true power spectrum is given by, 
\beq
W^{AMTM}(\bk) = \frac{\sum_{\alpha_1,\alpha_2} b^2_{(\alpha_1\alpha_2)}(\bk) {\tilde V}^{(\alpha_1\alpha_2;N_{1} N_{2})}(\bk)}{\sum_{\alpha_1, \alpha_2}b^2_{(\alpha_1\alpha_2)}(\bk)}.
\eeq 
One is obviously interested in the angle averaged pseudo power spectrum,
\beq
\label{angularAvPseudo}
\hat P^{W}(k) = \int \frac{d\theta}{(2\pi)} P^{W}(\bk)
\eeq
and as demonstrated in appendix \ref{appendixA}, this can be expressed in terms of the true power spectrum as,
\beq
\av{ \hat P^{W}(k) }=  \int dk' M(k,k^{'})  P(k')
\eeq
where ${M}$ is the mode-mode coupling kernel and depends on the power spectrum of the window $W$. After the power spectrum is binned, the above relation can be conveniently expressed as a matrix multiplication (see appendix  \ref{appendixA},),
\beq
\av{ \hat P^{W}_{b} }= \sum_{b'}{\tilde{M}}_{bb'}P_{b'} 
\eeq
where $b$ are the indices for the bins and $\tilde{\mathbb M}$ is the binned mode -mode coupling matrix. The binned mode-mode coupling matrix is more stable to inversion than the un-binned one which tends to be nearly singular, and therefore an estimate of the true power spectrum can be obtained by inverting the above relation,
\beq
\label{deConvol}
\hat P_{b} = (\tilde{\mathbb{M}}^{-1})_{bb'}\hat P^{W}_{b'}.
\eeq
\begin{figure}
\includegraphics[scale=0.38]{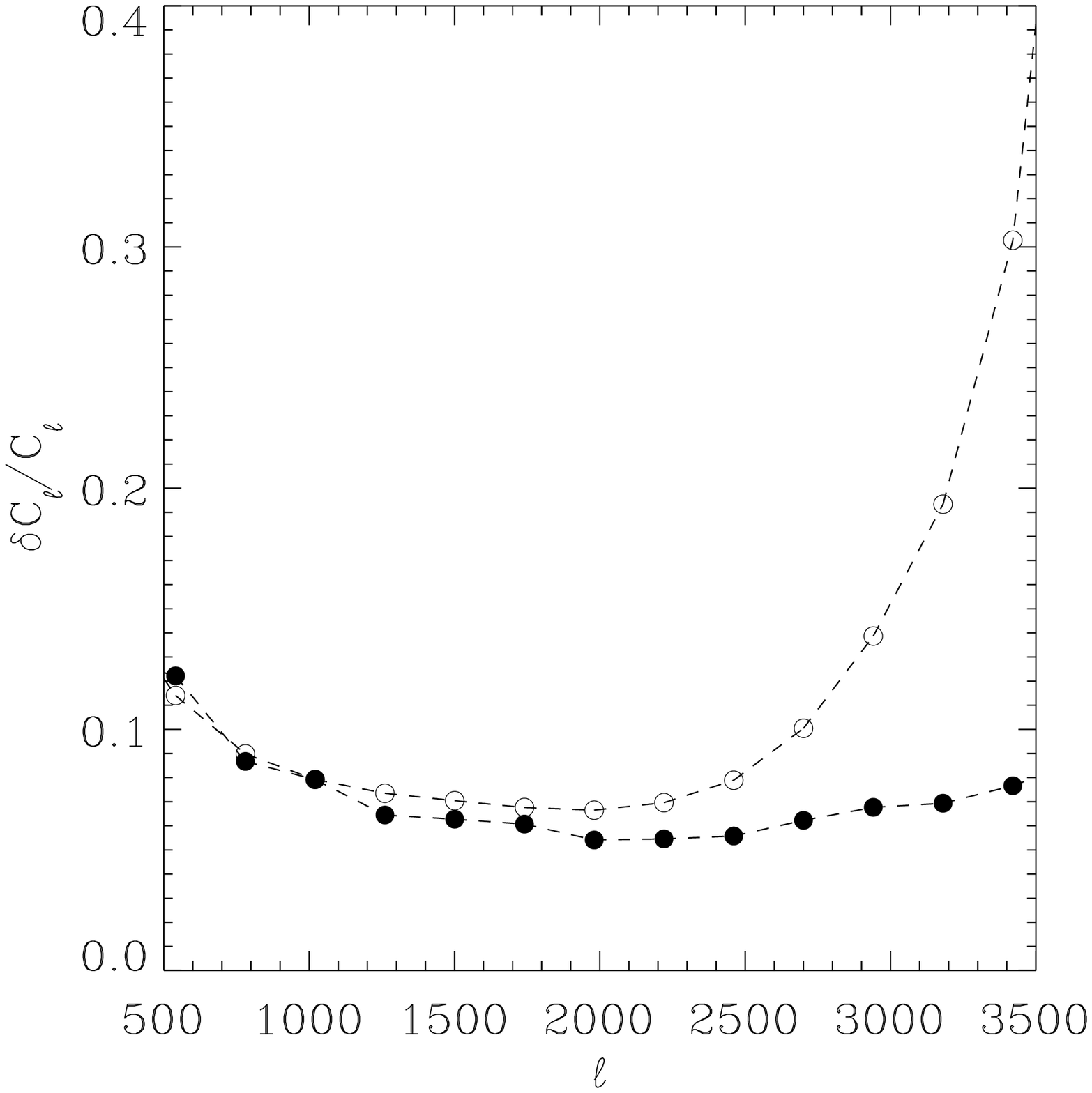}
\caption{\label{errorCompare} Fractional errors in the deconvolved binned power spectrum. The open circles represent the fractional errors for the periodogram method (left panel of Fig.~\ref{slavesAndMasters}). The filled circles represent the same for the AMTM method ( right panel of Fig.~\ref{slavesAndMasters}). Although the deconvolved power spectrum obtained from either method is an unbiased estimate of the true power spectrum, the errors from the periodogram method are much larger at high $\ell$ because of the highly biased nature of the mode-coupled periodogram at those multipoles. }
\end{figure}
We first illustrate the deconvolution of the power spectrum without a point source mask. For this purpose, we simulate a large $16$ degree square map, with $768$ pixels on a side, using a theoretical power spectrum. We then perform power spectrum analysis of the central 8 degree square portion of it. For the periodogram, we simply zero out the portion of the larger map outside the central 8 degree square, so that the mask becomes a zero-padded top-hat. For the AMTM, we discard the outer regions and  perform the adaptive multitaper analysis with $\nres=3$, $\ntap=5^{2}$  on the central 8 degree square. We generate the mode coupled power spectrum with each method and then deconvolve it using \eqref{deConvol}.   The results from $800$ Monte Carlo experiments are shown in Fig.~\ref{slavesAndMasters}. \par 
Firstly, we note that the construction of the mode-mode coupling matrix, as delineated in appendix~\ref{appendixA}, enables us to accurately predict the mode-mode coupled theoretical power spectrum. As such,  likelihood analyses for cosmological parameters can potentially be performed with the mode-coupled AMTM power spectrum, and this is specially appealing because the statistical properties of the mode-coupled binned power spectra can be precisely predicted (see \S~\ref{statProp}). 
\par Secondly, we find that the deconvolution of the periodogram as well as the AMTM power spectrum produces an unbiased estimate of the input power spectrum, but with a very important difference: the uncertainty in the deconvolved periodogram at $\ell\gtrsim 1500$ is larger than that in the deconvolved AMTM spectrum, and the difference continues to grow to factors of several for higher multipoles. This is seen clearly in Fig.~\ref{errorCompare}, where we have plotted the fractional errors in the deconvolved binned power spectrum   against the bin centers. For instance, at $\ell\sim 3000$ the periodogram method produces $\sim 3$ times larger error bars. This owes to the fact that due to spectral leakage, the periodogram produces a pseudo power spectrum that is highly biased relative to the input theory over these higher order multipoles. This bias adds to the uncertainty in the deconvolved power spectrum.  On the other hand, AMTM produces a nearly unbiased, albeit smoothed, pseudo power spectrum which, when deconvolved, does not incur any excess uncertainties. Ignoring point source masks for the moment, which are very specific to the CMB, this result alone immediately elevates the multitaper method to a far superior status than the periodogram, as a power spectrum estimation method for highly colored spectrum measured from finite maps. \par
One must bear in mind, that over the range of multipoles where leakage is not a serious problem, and the periodogram is essentially unbiased, the errors bars obtained from the periodogram are the smallest. This is because the periodogram makes use of all the modes that are available from the entire map, while each taper in a MTM process makes use of a certain fraction of the map. By using several tapers, one can compete with the periodogram error bars over the nearly white part of the spectrum. This is essentially seen in the convergence of the fractional error bars from the two methods in the low $\ell$ portion of  Fig.~\ref{errorCompare}. This, however, is not in contradiction to the fact, as shown in \S~\ref{statProp} that it is possible to have lower uncertainties in any bin for the pseudo power spectrum with AMTM than with the periodogram, by choosing a high enough $\nres$ (and consequently high $\ntap$). Unfortunately, this advantage goes away when the power spectrum is deconvolved.\par
\begin{figure}
\includegraphics[scale=0.38]{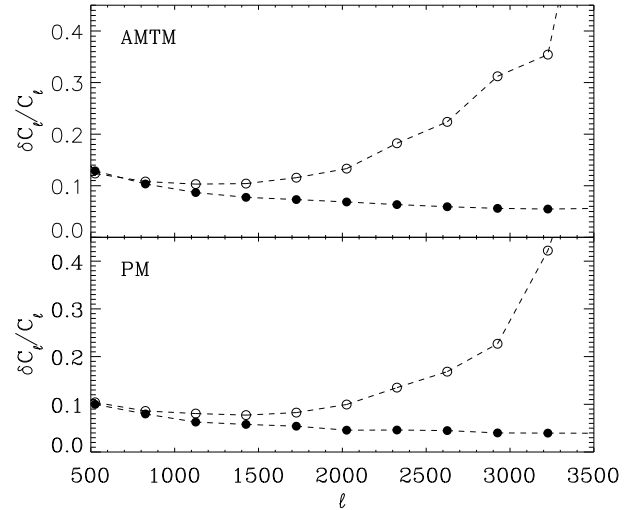}
\caption{\label{PWMasteredErrors} Effect of prewhitening on the errors in the deconvolved power spectrum in presence of a point source mask. \emph{Top Panel:} For the AMTM method: The open circles represent the fractional $1\sigma$ error bars on the deconvolved power spectrum when prewhitening is not performed, showing that holes in the point source mask render the AMTM method biased at high multipoles and lead to large error bars. The filled circles represent the same after prewhitening has been performed, showing that prewhitening remedies the leakage of power and makes the power spectrum estimator nearly unbiased. {\emph{Bottom Panel:}} The same as above, but for the periodogram or straight FFT. Note that prewhitening, if performed properly, makes the periodogram as good a power spectrum tool as the AMTM.}
\end{figure}

Next we turn to the practical issue of dealing with a point source mask for  a CMB map. As already discussed in \S~\ref{prewhiten}, the holes in the mask couple power over various scales, and neither the periodogram nor the AMTM remain an unbiased method at large multipoles, necessitating the prewhitening of the map. To study the effect of the mask and prewhitening on the deconvolved power spectrum, we multiply our map with a point source mask and repeat the Monte Carlo exercise with and without prewhitening.  The mask has a covering fraction of $98\%$ and the holes are of random sizes. For the prewhitening operation we use the parameters $R=1^{\prime}$ and $\alpha=0.02$.  The results on the fractional errors for this case are displayed in Fig.~\ref{PWMasteredErrors}. This figure shows the power of prewhitening approach. Prewhitening significantly reduces the errors. It is obvious that like the periodogram, AMTM is defenseless against the point source holes in the mask. This is expected because a taper with holes no longer retains the nice property of being a band-limited function. For both the top-hat  and the effective tapered window, the  sidelobes become dominated by the power produced by the holes. In such a situation, prewhitening the map becomes a necessity to reduce the uncertainties in the deconvolved power spectrum. With proper prewhitening both the periodogram and the AMTM method can produce nearly unbiased pseudo power spectra and consequently small error bars at the high multipoles, as evident from Fig.~\ref{PWMasteredErrors}. It is important to note here that the power spectrum of a top-hat has greater sidelobe power than the effective taper in the intermediate range of multipoles between the regions where the central lobe is falling and where the point source power starts to dominate. Therefore, depending on the quality of prewhitening that can be performed on a map, AMTM may be a more reliable method to apply on the prewhitened and masked map, rather than the periodogram. 

\section{Conclusion \label{conclusion}}
Power spectrum estimation on small sections of the CMB sky is a non-trivial problem due to spectral leakage from the finite nature of the patch, which is further compounded by the application of point source masks. The direct application of standard decorrelation techniques,  like the MASTER algorithm \citep{2002:Hivon}, to obtain an unbiased estimate of the power spectrum leads to unnecessarily large uncertainties at high multipoles due to the highly biased nature of the pseudo power spectrum at those multipoles. We have put forward two techniques to reduce the uncertainties in the deconvolved power spectrum. First, we have formulated a two-dimensional adaptive multitaper method (AMTM) which produces nearly unbiased pseudo power spectra for maps without point source masks, by minimizing the leakage of power due to the finite size of the patch. This is achieved at the cost of lowered spectral resolution. The deconvolution of the pseudo power spectrum so produced, leads to an unbiased estimate of the true power spectrum that has several times smaller error bars at high multipoles than the deconvolved periodogram. In presence of point source masks, however, this method becomes non-optimal because the pseudo power spectrum estimated even by AMTM is no longer unbiased. To deal with the point source mask, we have put forward a novel way of prewhitening a CMB map, with manifestly local operations which has simple representations in the Fourier space. This operation produces a map, the power spectrum of which has several orders of magnitude lower dynamic range than the original map. This renders the leakage of power due to holes and edges a relatively benign issue for the prewhitened map. If the prewhitening operation can be tuned to make the power spectrum of the map nearly white, a pseudo power spectrum obtained via a simple periodogram may be nearly unbiased and therefore, can be deconvolved to give a precisely unbiased estimate of the power spectrum, thereby avoiding unnecessarily large error bars at large multipoles. If the map cannot be made sufficiently white for a periodogram, an AMTM method can be applied to the prewhitened map to guard against leakage and achieve the same result. We have shown that by applying these methods, one can reduce the error bar in the small scale power spectrum by a factor of $\sim 4$ at $\ell\sim 3500$. This can be translated into a many-fold reduction in the required integration time of a CMB experiment to achieve some target uncertainty on the small scale power spectrum than that dictated by standard techniques.
\appendix
\begin{widetext}
\section{\label{appendixZ}Statistical properties of multitaper estimators}
The statistical properties of the multitaper spectral estimators follow from the basic result \eqref{periodogramPDF} for the distribution of the periodogram. In the following, we will use the equivalent degrees of freedom argument to derive the distribution functions for the MTM estimators. The first result we will use is that the probability distribution of an eigenspectrum $\hat P^{(\alpha_{1}\alpha_{2})}(\bk)$ at each frequency-space point $\bk$ (i.e. at each pixel)  will have the same form as  \eqref{periodogramPDF} for asymptotically large $N$. Since the final spectral estimate is a weighted sum of these eigenspectra, it is reasonable to assume that the former is distributed also as a scaled $\chi^{2}$ variable. Let us hypothesize,
\beq
P^{\mathrm{AMTM}}(\bk) \eqd a \chi^{2}_{\nu}
\eeq
where both $\nu$ and $a$ are unknown. Now, we will make use of the facts that $\av{P^{\mathrm{AMTM}}} = \av{a\chi^{2}_{\nu}} = a\nu$ and $\mathrm{var}(P^{\mathrm{AMTM}}) =\mathrm{var}(a\chi^{2}_{\nu}) = 2 a^{2}\nu $. Therefore,
\beq
\nu = \frac{2\av{P^{\mathrm{AMTM}}}^{2}}{ \mathrm{var}(P^{\mathrm{AMTM}})} \: ; \: \: a = \frac{\av{P^{\mathrm{AMTM}}}}{\nu}.
\eeq
Since the eigenspectra $\hat P^{\alpha_{1}\alpha_{2}}$ are asymptotically uncorrelated and are also unbiased estimators of $P(\bk)$, we obtain, using the definition of the AMTM estimator \eqref{AMTM2D},
\beqn
\av{P^{\mathrm{AMTM}}} & = &  P(\bk), \\
\mathrm{var}(P^{\mathrm{AMTM}}) & =  & P^{2}(\bk) \frac{\sum_{\alpha_1,\alpha_2} b^4_{(\alpha_1\alpha_2)}(\bk)}{\left(\sum_{\alpha_1, \alpha_2}b^2_{(\alpha_1\alpha_2)}(\bk)\right)^{2}}
\eeqn
for $N \rightarrow \infty$, which immediately gives us
\beq
\nu(\bk) = \frac{2\left(\sum_{\alpha_1, \alpha_2}b^2_{(\alpha_1\alpha_2)}(\bk)\right)^{2}}{\sum_{\alpha_1,\alpha_2} b^4_{(\alpha_1\alpha_2)}(\bk)}.
\eeq
which is equation \eqref{amtmNu}.
\par
Now we turn to the approximate form of the distribution for the binned power spectrum, $P^{B}$ \eqref{binnedPower}. We again postulate that $P^{B}$ is distributed as $c \chi^{2}_{\nu_{b}}$, so that,
 \beq
 \label{nuAndC}
 \nu_{b} = \frac{2\av{P^{B}}^{2}}{ \mathrm{var}(P^{B})} \: ; \: \: c = \frac{\av{P^{\mathrm{B}}}}{\nu_{b}}.		
 \eeq  
 Now note that because the images or maps we will be concerned with are real, only the half-plane in the $\bk$ space are independent because of reflection symmetry i.e $P(\bk) = P(-\bk)$. For the time being, let us assume that all $\bk$ space pixels are uncorrelated, so that all pixels in one half of the $\bk$ plane are independent of each other. Later, we will correct this for the fact the tapers highly correlate adjacent pixels in each resolution element. With these assumptions, \eqref{binnedPower} can be re-written as,
 \beq
 P^{B}(k_{b}) = \frac{2}{N_{b}}\sum_{i,j \in b; k_{j}>0} P^{\mathrm{AMTM}}(k_{i},k_{j}).
 \eeq
 Therefore, the variance is given by,
 \beqn
 \mathrm{var}(P^{B}(k_{b}) ) &=& \frac{4}{N_{b}^{2}} \sum_{i,j \in b; k_{j}>0} \frac{2}{\nu(k_{i},k_{j})} \av{P^{\mathrm{AMTM}}(k_{i},k_{j})}^{2}\nn\\
 &=& \frac{4}{N_{b}^{2}} \av{P^{\mathrm{AMTM}}(\modu{\bk}=k_{b})}^{2}\nn\\
 && \times\sum_{i,j \in b} \frac{1}{\nu(k_{i},k_{j})} 
 \eeqn
 where in the last step we have assumed that the power spectrum is slowly varying inside a bin. The factor of $2$ inside the summation goes away because we have extended the summation to include the lower half plane. Using \eqref{nuAndC}, we therefore get
 \beq
 \frac 1{\nu_{b}} = \frac2 {N_{b}^{2}} \sum_{i,j \in b} \frac{1}{\nu(k_{i},k_{j})}.
 \eeq
Now, we will correct for the fact that all pixels in the upper half plane are not independent. In fact for a taper with a given $N_{res}$, approximately $N_{res}^2$ pixels get highly correlated. Therefore we can think of ``super-pixels'' of dimension  $N_{res} \times N_{res}$, which are independent of each other. If we repeat the calculation above with this assumption, then  we obtain the result,
\beq
 \frac 1{\nu_{b}} \simeq \frac{2 N_{res}^2} {N_{b}^{2}} \sum_{i,j \in b} \frac{1}{\nu(k_{i},k_{j})},
\eeq 
which essentially reduces the degrees-of-freedom by the number of pixels that fall in each super-pixel. This is equation \eqref{binnedPDFNu}.\par

\section{\label{appendixA} Mode-mode coupling matrix}
In the following, we derive the form of the mode-mode coupling kernel. For the sake of completeness, we repeat and expand upon some of the calculations from appendix A.1 of \cite{2002:Hivon}.  Note that the symbol $k$ (frequency) in the said appendix of that paper is smaller by a factor of $2\pi$ than the $k$ (angular frequency) used in the present section. In other words, the $k$ used here is readily identified with the multipole $\ell$ in the flat-sky approximation, while it would stand for $\ell/(2\pi)$ in the aforementioned appendix.  \par
We start with the definition \eqref{angularAvPseudo} of the angle averaged pseudo power spectrum and with the aid of \eqref{pseudocl} express it as,
\beq
\label{pw}
\av{\hat P^{W}(k_{1})} = \int \frac{d\theta_{1}}{(2\pi)}\int\frac{ d^{2} {\bk_{2}}}{(2 \pi)^{2}} \frac{{\modu{W(\bk_{1}-\bk_{2})}}^{2}}{(2\pi)^{2}} P(k_{2}).
\eeq
We can express the Fourier mode of the window appearing above as
\beq
W(\bk_{1}-\bk_{2}) = \int d^{2}\bk_{3} W(\bk_{3}) \delta^{2}(\bk_{3}-\bk_{1}+\bk_{2})
\eeq
which immediately gives,
\beq
\frac{{\modu{W(\bk_{1}-\bk_{2})}}^{2}}{(2\pi)^{2}}  = 2\pi \int dk_{3}~ k_{3} {\cal W}(k_{3}) \delta^{2}(\bk_{3}-\bk_{1}+\bk_{2})
\eeq
where we have introduced the power spectrum of the window,
\beq
 (2\pi)^{2}{\cal W}(k) = \int \frac{d\theta}{(2\pi)}   W^{*}(\bk) W(\bk).
\eeq
Substituting in \eqref{pw}, we get,
\beqn
\label{pw1}
\av{\hat P^{W}(k_{1})} &=& \int \frac{d\theta_{1}}{(2\pi)}\int\frac{ dk_{2}~ k_{2}}{(2 \pi)}\int d\theta_{2} P(k_{2})\int dk_{3}~ k_{3} {\cal W}(k_{3}) \delta^{2}(\bk_{3}-\bk_{1}+\bk_{2})\\
&=& \int dk_{2} M_{k_{1}k_{2}} P(k_{2}),
\eeqn
where we have introduced the mode-mode coupling matrix,
\beq
\label{mcm}
M_{k_{1}k_{2}}  = \frac{k_{2}}{(2\pi)}\int dk_{3}~ k_{3} {\cal W}(k_{3}) J(k_{1},k_{2},k_{3})
\eeq
with,
\beq
J(k_{1},k_{2},k_{3}) =  \int \frac{d\theta_{1}}{(2\pi)} \int d\theta_{2} \delta^{2}(\bk_{3}-\bk_{1}+\bk_{2})
\eeq
It is instructive to compare \eqref{mcm} with the full-sky result from appendix A.2 of \cite{2002:Hivon}, 
\beq
\label{mcml}
M_{\ell_{1}\ell_{2}} = \frac{2\ell_{2}+1}{4\pi} \sum_{\ell_{3}} (2\ell_{3}+1) {\cal W}_{\ell_{3}} 
\left(
\begin{array}{ccc}
\ell_{1} & \ell_{2} & \ell_{3}\\
0 & 0 & 0
\end{array}\right)^{2}.
\eeq
It can be shown that for large $\ell$, 
\beq
\left(
\begin{array}{ccc}
\ell_{1} & \ell_{2} & \ell_{3}\\
0 & 0 & 0
\end{array}\right)^{2} \rightarrow J(\ell_{1},\ell_{2},\ell_{3})
\eeq
so that if we identify $k_{i}$ with $\ell_{i}$, the large $\ell$ limit of the spherical mode-mode coupling matrix goes correctly to the flat-sky expression \eqref{mcm}. Note that the appearance of an extra factor of $2$ in the large $\ell$ limit of the summand in \eqref{mcml} over the integrand in \eqref{mcm} is fine because the sum is restricted to values $\ell_{3}$ which make $(\ell_{1}+\ell_{2}+\ell_{3})$ even, while the integral over $k$ spans the entire allowed range for each $(k_{1},k_{2})$  pair.\par
We now turn to the evaluation of the $J$ function. Using the factorized form of the delta function in plane polar coordinates,
$\delta(\bmm r - \bmm r') ={\delta(r-r')}\delta(\theta-\theta')/r$ ,
we can write,
\beq
J(k_{1},k_{2},k_{3}) =  \frac 1{(2\pi)} \int d\theta_{2}\delta^{2}(k_{1}-\modu{\bk_{3}+\bk_{2}})/k_{1}.
\eeq
The integral over $\theta_{2}$ can be performed by using the following property of the delta function,
\beq
\delta(g(x)) = \sum_{i} \frac{\delta(x-x_{i})}{\modu{g'(x_{i})}}
\eeq
where $x_{i}$ are the roots of $g(x)=0$. In our case there are two roots and the integral finally yields
\beq
J(k_{1},k_{2},k_{3}) =  \frac 2 {\pi}  \frac 1 {\sqrt{-K(K-2k_{1})(K-2k_{2})(K-2k_{3})}}
\eeq
 for $\modu{k_{2}-k_{3}}<k_{1}<k_{1}+k_{2}$, and zero outside the interval, where $K = k_{1} + k_{2} + k_{3}$. As a word of caution against using this form directly into the integral \eqref{mcm}, we would like to point out that due to the diverging nature of the integrand near the edges of the allowed interval, numerical integration schemes that span the entire range will be unreliable. Instead we find, drawing parallels with the spherical harmonic case, that  approximating the integration in \eqref{mcm} by a sum over integral values of $k$, by interpolating ${\cal W}(k)$ onto integers, provides a very stable and reliable way of computing the mode coupling matrix. In fact, the boundaries of interval can be included in the sum by using the large $\ell$ limit of the $3-j$ functions at those points,
\beq
\left(
\begin{array}{ccc}
\ell_{1} & \ell_{2} & \ell_{3}\\
0 & 0 & 0
\end{array}\right)^{2} \rightarrow 
\begin{cases}
   \frac 1 {2\sqrt{\pi \ell_{1} \ell_{2} \ell_{3}}} & \text{for $\ell_{3} \rightarrow \ell_{1}+\ell_{2}$ or $\ell_{3}\rightarrow  \modu{\ell_{1}-\ell_{2}}$} \\
   \frac{\delta_{\ell_{1}\ell_{2}} }{2 \ell_{1}} & \text{for $\ell_{3}=0$}
\end{cases}
.
\eeq
At the end of the exercise, we therefore generate the mode-coupling matrix at integer subscripts, which by abusing notation a bit, we will denote by $M_{\ell \ell'}$. \par
Next, we consider the binning of the pseudo power spectrum. In order to construct the mode-mode coupling kernel relevant to a binned power spectrum, we first define the binning operator $B$, which performs the binning of the integral indexed quantities to binned values and its reciprocal operation $U$. Given the lower and upper boundaries,$\ell^{b}_{\mathrm{low}}$ and $\ell^{b}_{\mathrm{high}}$ of a bin $b$, a simple form of these operators can be written as,
\beq
B_{b\ell} = 
\begin{cases}
\frac{\ell^{\alpha}}{(\ell^{b}_{\mathrm{high}}-\ell^{b}_{\mathrm{low}})}, & \text{if $2\le \ell^{b}_{\mathrm{low}}\le \ell < \ell^{b}_{\mathrm{high}}$}\\
0, & \text{otherwise,}
\end{cases}
\eeq
and 
\beq
U_{\ell b} =
\begin{cases}
 \frac 1{\ell^{\alpha}}, & \text{if $2\le \ell^{b}_{\mathrm{low}}\le \ell < \ell^{b}_{\mathrm{high}}$}\\
0, & \text{otherwise.}
\end{cases}
\eeq
Here, $\alpha$ is chosen to make the power spectrum ``flatter'' and for the damping tail of the CMB, a suitable value is $\alpha = 4$. \par
Note that the pseudo power spectrum realized on the two-dimensional $\bk$-space may be directly binned into the bins $b$, by averaging the value in the pixels that fall inside the annuli demarcated by the bin boundaries, without having to go through an intermediate step of interpolating the power spectrum onto integers. The binned pseudo power spectrum is, therefore related to the binned true power spectrum as
\beqn
\nn \av{P^{W}_{b}} &=& B_{b\ell} M_{\ell \ell'} P_{\ell'}\\
\nn & = &  B_{b\ell} M_{\ell \ell'} U_{\ell' b'} B_{b' \ell'} P_{\ell'}\\
& = & {\tilde M}_{bb'} P_{b'},
\eeqn
where we have defined the binned mode-mode coupling matrix as $\tilde{\mathbb{M}} = {\mathbb{B}}^{T} \mathbb M \mathbb U$.

\end{widetext}
\acknowledgements
We dedicate this paper to the memory of eminent geophysicist  F. Anthony Dahlen (1942-2007), who had suggested to us the idea of multitapering for the CMB. We would like to thank Frederick J. Simons, Mark A. Wieczorek, Neelima Sehgal and Tobias A. Marriage for enlightening discussions and Viviana Acquaviva for useful comments on the manuscript. SD would like to acknowledge the warm hospitality extended by the mentors and colleagues at Jadwin hall, specially Lyman Page and Jo Dunkley, while the Department of Astrophysics was under renovation. AH would like to thank Eric Hivon for useful discussions. AH is supported by the LTSA program. SD is supported by the  Charlotte Elizabeth Procter Honorific Fellowship from Princeton University and  NSF  grant 0707731. DNS acknowledges support from NASA ATP grant NNX 08AH30G.

\begin{thebibliography}{0}
\expandafter\ifx\csname natexlab\endcsname\relax\def\natexlab#1{#1}\fi
\expandafter\ifx\csname bibnamefont\endcsname\relax
  \def\bibnamefont#1{#1}\fi
\expandafter\ifx\csname bibfnamefont\endcsname\relax
  \def\bibfnamefont#1{#1}\fi
\expandafter\ifx\csname citenamefont\endcsname\relax
  \def\citenamefont#1{#1}\fi
\expandafter\ifx\csname url\endcsname\relax
  \def\url#1{\texttt{#1}}\fi
\expandafter\ifx\csname urlprefix\endcsname\relax\def\urlprefix{URL }\fi
\providecommand{\bibinfo}[2]{#2}
\providecommand{\eprint}[2][]{\url{#2}}

\end{thebibliography}


\begin{thebibliography}{31}
\expandafter\ifx\csname natexlab\endcsname\relax\def\natexlab#1{#1}\fi
\expandafter\ifx\csname bibnamefont\endcsname\relax
  \def\bibnamefont#1{#1}\fi
\expandafter\ifx\csname bibfnamefont\endcsname\relax
  \def\bibfnamefont#1{#1}\fi
\expandafter\ifx\csname citenamefont\endcsname\relax
  \def\citenamefont#1{#1}\fi
\expandafter\ifx\csname url\endcsname\relax
  \def\url#1{\texttt{#1}}\fi
\expandafter\ifx\csname urlprefix\endcsname\relax\def\urlprefix{URL }\fi
\providecommand{\bibinfo}[2]{#2}
\providecommand{\eprint}[2][]{\url{#2}}

\bibitem[{\citenamefont{{Hajian} and {Souradeep}}(2006)}]{HS06}
\bibinfo{author}{\bibfnamefont{A.}~\bibnamefont{{Hajian}}} \bibnamefont{and}
  \bibinfo{author}{\bibfnamefont{T.}~\bibnamefont{{Souradeep}}},
  \bibinfo{journal}{\prd} \textbf{\bibinfo{volume}{74}},
  \bibinfo{pages}{123521} (\bibinfo{year}{2006}),
  \eprint{arXiv:astro-ph/0607153}.

\bibitem[{\citenamefont{{Komatsu} et~al.}(2003)\citenamefont{{Komatsu},
  {Kogut}, {Nolta}, {Bennett}, {Halpern}, {Hinshaw}, {Jarosik}, {Limon},
  {Meyer}, {Page} et~al.}}]{Komatsu}
\bibinfo{author}{\bibfnamefont{E.}~\bibnamefont{{Komatsu}}},
  \bibinfo{author}{\bibfnamefont{A.}~\bibnamefont{{Kogut}}},
  \bibinfo{author}{\bibfnamefont{M.~R.} \bibnamefont{{Nolta}}},
  \bibinfo{author}{\bibfnamefont{C.~L.} \bibnamefont{{Bennett}}},
  \bibinfo{author}{\bibfnamefont{M.}~\bibnamefont{{Halpern}}},
  \bibinfo{author}{\bibfnamefont{G.}~\bibnamefont{{Hinshaw}}},
  \bibinfo{author}{\bibfnamefont{N.}~\bibnamefont{{Jarosik}}},
  \bibinfo{author}{\bibfnamefont{M.}~\bibnamefont{{Limon}}},
  \bibinfo{author}{\bibfnamefont{S.~S.} \bibnamefont{{Meyer}}},
  \bibinfo{author}{\bibfnamefont{L.}~\bibnamefont{{Page}}},
  \bibnamefont{et~al.}, \bibinfo{journal}{Apjs} \textbf{\bibinfo{volume}{148}},
  \bibinfo{pages}{119} (\bibinfo{year}{2003}), \eprint{arXiv:astro-ph/0302223}.

\bibitem[{\citenamefont{{Gorski} et~al.}(1996)\citenamefont{{Gorski}, {Banday},
  {Bennett}, {Hinshaw}, {Kogut}, {Smoot}, and {Wright}}}]{COBE}
\bibinfo{author}{\bibfnamefont{K.~M.} \bibnamefont{{Gorski}}},
  \bibinfo{author}{\bibfnamefont{A.~J.} \bibnamefont{{Banday}}},
  \bibinfo{author}{\bibfnamefont{C.~L.} \bibnamefont{{Bennett}}},
  \bibinfo{author}{\bibfnamefont{G.}~\bibnamefont{{Hinshaw}}},
  \bibinfo{author}{\bibfnamefont{A.}~\bibnamefont{{Kogut}}},
  \bibinfo{author}{\bibfnamefont{G.~F.} \bibnamefont{{Smoot}}},
  \bibnamefont{and} \bibinfo{author}{\bibfnamefont{E.~L.}
  \bibnamefont{{Wright}}}, \bibinfo{journal}{\apjl}
  \textbf{\bibinfo{volume}{464}}, \bibinfo{pages}{L11+} (\bibinfo{year}{1996}),
  \eprint{arXiv:astro-ph/9601063}.

\bibitem[{\citenamefont{{Nolta} et~al.}(2008)\citenamefont{{Nolta}, {Dunkley},
  {Hill}, {Hinshaw}, {Komatsu}, {Larson}, {Page}, {Spergel}, {Bennett}, {Gold}
  et~al.}}]{Nolta}
\bibinfo{author}{\bibfnamefont{M.~R.} \bibnamefont{{Nolta}}},
  \bibinfo{author}{\bibfnamefont{J.}~\bibnamefont{{Dunkley}}},
  \bibinfo{author}{\bibfnamefont{R.~S.} \bibnamefont{{Hill}}},
  \bibinfo{author}{\bibfnamefont{G.}~\bibnamefont{{Hinshaw}}},
  \bibinfo{author}{\bibfnamefont{E.}~\bibnamefont{{Komatsu}}},
  \bibinfo{author}{\bibfnamefont{D.}~\bibnamefont{{Larson}}},
  \bibinfo{author}{\bibfnamefont{L.}~\bibnamefont{{Page}}},
  \bibinfo{author}{\bibfnamefont{D.~N.} \bibnamefont{{Spergel}}},
  \bibinfo{author}{\bibfnamefont{C.~L.} \bibnamefont{{Bennett}}},
  \bibinfo{author}{\bibfnamefont{B.}~\bibnamefont{{Gold}}},
  \bibnamefont{et~al.}, \bibinfo{journal}{ArXiv e-prints}
  \textbf{\bibinfo{volume}{803}} (\bibinfo{year}{2008}), \eprint{0803.0593}.

\bibitem[{\citenamefont{{Hinshaw} et~al.}(2008)\citenamefont{{Hinshaw},
  {Weiland}, {Hill}, {Odegard}, {Larson}, {Bennett}, {Dunkley}, {Gold},
  {Greason}, {Jarosik} et~al.}}]{Hinshaw}
\bibinfo{author}{\bibfnamefont{G.}~\bibnamefont{{Hinshaw}}},
  \bibinfo{author}{\bibfnamefont{J.~L.} \bibnamefont{{Weiland}}},
  \bibinfo{author}{\bibfnamefont{R.~S.} \bibnamefont{{Hill}}},
  \bibinfo{author}{\bibfnamefont{N.}~\bibnamefont{{Odegard}}},
  \bibinfo{author}{\bibfnamefont{D.}~\bibnamefont{{Larson}}},
  \bibinfo{author}{\bibfnamefont{C.~L.} \bibnamefont{{Bennett}}},
  \bibinfo{author}{\bibfnamefont{J.}~\bibnamefont{{Dunkley}}},
  \bibinfo{author}{\bibfnamefont{B.}~\bibnamefont{{Gold}}},
  \bibinfo{author}{\bibfnamefont{M.~R.} \bibnamefont{{Greason}}},
  \bibinfo{author}{\bibfnamefont{N.}~\bibnamefont{{Jarosik}}},
  \bibnamefont{et~al.}, \bibinfo{journal}{ArXiv e-prints}
  \textbf{\bibinfo{volume}{803}} (\bibinfo{year}{2008}), \eprint{0803.0732}.

\bibitem[{\citenamefont{Reichardt et~al.}(2008)}]{ACBAR}
\bibinfo{author}{\bibfnamefont{C.~L.} \bibnamefont{Reichardt}}
  \bibnamefont{et~al.} (\bibinfo{year}{2008}), \eprint{0801.1491}.

\bibitem[{\citenamefont{{QUaD collaboration: C.~Pryke}
  et~al.}(2008)\citenamefont{{QUaD collaboration: C.~Pryke}, {Ade}, {Bock},
  {Bowden}, {Brown}, {Cahill}, {Castro}, {Church}, {Culverhouse}, {Friedman}
  et~al.}}]{QUAD}
\bibinfo{author}{\bibnamefont{{QUaD collaboration: C.~Pryke}}},
  \bibinfo{author}{\bibfnamefont{P.}~\bibnamefont{{Ade}}},
  \bibinfo{author}{\bibfnamefont{J.}~\bibnamefont{{Bock}}},
  \bibinfo{author}{\bibfnamefont{M.}~\bibnamefont{{Bowden}}},
  \bibinfo{author}{\bibfnamefont{M.~L.} \bibnamefont{{Brown}}},
  \bibinfo{author}{\bibfnamefont{G.}~\bibnamefont{{Cahill}}},
  \bibinfo{author}{\bibfnamefont{P.~G.} \bibnamefont{{Castro}}},
  \bibinfo{author}{\bibfnamefont{S.}~\bibnamefont{{Church}}},
  \bibinfo{author}{\bibfnamefont{T.}~\bibnamefont{{Culverhouse}}},
  \bibinfo{author}{\bibfnamefont{R.}~\bibnamefont{{Friedman}}},
  \bibnamefont{et~al.}, \bibinfo{journal}{ArXiv e-prints}
  \textbf{\bibinfo{volume}{805}} (\bibinfo{year}{2008}), \eprint{0805.1944}.

\bibitem[{ACT()}]{ACT}
\bibinfo{howpublished}{\url{http://www.physics.princeton.edu/act/}}.

\bibitem[{SPT()}]{SPT}
\bibinfo{howpublished}{\url{http://pole.uchicago.edu}}.

\bibitem[{Planck()}]{Planck}
\bibinfo{howpublished}{\url{http://www.rssd.esa.int/Planck/}}.

\bibitem[{\citenamefont{Percival and Walden}(1993)}]{Percival+93}
\bibinfo{author}{\bibfnamefont{D.~B.} \bibnamefont{Percival}} \bibnamefont{and}
  \bibinfo{author}{\bibfnamefont{A.~T.} \bibnamefont{Walden}},
  \emph{\bibinfo{title}{Spectral Analysis {f}or Physical Applications,
  Multitaper {a}nd Conventional Univariate Techniques}}
  (\bibinfo{publisher}{Cambridge Univ.~Press}, \bibinfo{address}{New York},
  \bibinfo{year}{1993}).

\bibitem[{\citenamefont{Tegmark}(1997)}]{Tegmark}
\bibinfo{author}{\bibfnamefont{M.}~\bibnamefont{Tegmark}},
  \bibinfo{journal}{Phys. Rev. D} \textbf{\bibinfo{volume}{56}},
  \bibinfo{pages}{4514} (\bibinfo{year}{1997}).

\bibitem[{\citenamefont{{Dor{\'e}} et~al.}(2001)\citenamefont{{Dor{\'e}},
  {Teyssier}, {Bouchet}, {Vibert}, and {Prunet}}}]{Dore}
\bibinfo{author}{\bibfnamefont{O.}~\bibnamefont{{Dor{\'e}}}},
  \bibinfo{author}{\bibfnamefont{R.}~\bibnamefont{{Teyssier}}},
  \bibinfo{author}{\bibfnamefont{F.~R.} \bibnamefont{{Bouchet}}},
  \bibinfo{author}{\bibfnamefont{D.}~\bibnamefont{{Vibert}}}, \bibnamefont{and}
  \bibinfo{author}{\bibfnamefont{S.}~\bibnamefont{{Prunet}}},
  \bibinfo{journal}{\aap} \textbf{\bibinfo{volume}{374}}, \bibinfo{pages}{358}
  (\bibinfo{year}{2001}), \eprint{arXiv:astro-ph/0101112}.

\bibitem[{\citenamefont{{Smith} et~al.}(2007)\citenamefont{{Smith}, {Zahn}, and
  {Dor{\'e}}}}]{Smith}
\bibinfo{author}{\bibfnamefont{K.~M.} \bibnamefont{{Smith}}},
  \bibinfo{author}{\bibfnamefont{O.}~\bibnamefont{{Zahn}}}, \bibnamefont{and}
  \bibinfo{author}{\bibfnamefont{O.}~\bibnamefont{{Dor{\'e}}}},
  \bibinfo{journal}{\prd} \textbf{\bibinfo{volume}{76}},
  \bibinfo{pages}{043510} (\bibinfo{year}{2007}), \eprint{arXiv:0705.3980}.

\bibitem[{\citenamefont{{Hivon} et~al.}(2002)\citenamefont{{Hivon},
  {G{\'o}rski}, {Netterfield}, {Crill}, {Prunet}, and {Hansen}}}]{2002:Hivon}
\bibinfo{author}{\bibfnamefont{E.}~\bibnamefont{{Hivon}}},
  \bibinfo{author}{\bibfnamefont{K.~M.} \bibnamefont{{G{\'o}rski}}},
  \bibinfo{author}{\bibfnamefont{C.~B.} \bibnamefont{{Netterfield}}},
  \bibinfo{author}{\bibfnamefont{B.~P.} \bibnamefont{{Crill}}},
  \bibinfo{author}{\bibfnamefont{S.}~\bibnamefont{{Prunet}}}, \bibnamefont{and}
  \bibinfo{author}{\bibfnamefont{F.}~\bibnamefont{{Hansen}}},
  \bibinfo{journal}{\apj} \textbf{\bibinfo{volume}{567}}, \bibinfo{pages}{2}
  (\bibinfo{year}{2002}), \eprint{arXiv:astro-ph/0105302}.

\bibitem[{\citenamefont{Welch}(1967)}]{1161901}
\bibinfo{author}{\bibfnamefont{P.}~\bibnamefont{Welch}},
  \bibinfo{journal}{Audio and Electroacoustics, IEEE Transactions on}
  \textbf{\bibinfo{volume}{15}}, \bibinfo{pages}{70} (\bibinfo{year}{1967}),
  ISSN \bibinfo{issn}{0018-9278}.

\bibitem[{\citenamefont{Thomson}(1982)}]{Thomson:1982p3137}
\bibinfo{author}{\bibfnamefont{D.}~\bibnamefont{Thomson}},
  \bibinfo{journal}{Proceedings of the IEEE} \textbf{\bibinfo{volume}{70}},
  \bibinfo{pages}{1055} (\bibinfo{year}{1982}).

\bibitem[{\citenamefont{{Slepian}}(1978)}]{1978ATTTJ..57.1371S}
\bibinfo{author}{\bibfnamefont{D.}~\bibnamefont{{Slepian}}},
  \bibinfo{journal}{AT T Technical Journal} \textbf{\bibinfo{volume}{57}},
  \bibinfo{pages}{1371} (\bibinfo{year}{1978}).

\bibitem[{\citenamefont{Bell et~al.}(1993)\citenamefont{Bell, Percival, and
  Walden}}]{1993:Bell}
\bibinfo{author}{\bibfnamefont{B.}~\bibnamefont{Bell}},
  \bibinfo{author}{\bibfnamefont{D.~B.} \bibnamefont{Percival}},
  \bibnamefont{and} \bibinfo{author}{\bibfnamefont{A.~T.}
  \bibnamefont{Walden}}, \bibinfo{journal}{Journal of Computational and
  Graphical Statistics} \textbf{\bibinfo{volume}{2}}, \bibinfo{pages}{119}
  (\bibinfo{year}{1993}), ISSN \bibinfo{issn}{10618600},
  \urlprefix\url{http://www.jstor.org/stable/1390958}.

\bibitem[{\citenamefont{Cramer}(1940)}]{1940:Cramer}
\bibinfo{author}{\bibfnamefont{H.}~\bibnamefont{Cramer}}, \bibinfo{journal}{The
  Annals of Mathematics} \textbf{\bibinfo{volume}{41}}, \bibinfo{pages}{215}
  (\bibinfo{year}{1940}), ISSN \bibinfo{issn}{0003486X},
  \urlprefix\url{http://www.jstor.org/stable/1968827}.

\bibitem[{\citenamefont{Doob}(1963)}]{doob:172}
\bibinfo{author}{\bibfnamefont{J.~L.} \bibnamefont{Doob}},
  \bibinfo{journal}{SIAM Review} \textbf{\bibinfo{volume}{5}},
  \bibinfo{pages}{172} (\bibinfo{year}{1963}),
  \urlprefix\url{http://link.aip.org/link/?SIR/5/172/2}.

\bibitem[{\citenamefont{Priestly}(1988)}]{1988:Priestly}
\bibinfo{author}{\bibfnamefont{M.~B.} \bibnamefont{Priestly}},
  \bibinfo{journal}{Journal of the Royal Statistical Society. Series A
  (Statistics in Society)} \textbf{\bibinfo{volume}{151}}, \bibinfo{pages}{573}
  (\bibinfo{year}{1988}), ISSN \bibinfo{issn}{09641998},
  \urlprefix\url{http://www.jstor.org/stable/2983035}.

\bibitem[{\citenamefont{{{Barr, D. L.}} et~al.}(1988)\citenamefont{{{Barr, D.
  L.}}, {{Brown, W. L.}}, and {{Thompson, D. J.}}}}]{1988:Barr}
\bibinfo{author}{\bibnamefont{{{Barr, D. L.}}}},
  \bibinfo{author}{\bibnamefont{{{Brown, W. L.}}}}, \bibnamefont{and}
  \bibinfo{author}{\bibnamefont{{{Thompson, D. J.}}}}, \bibinfo{journal}{Le
  Journal de Physique Colloques} \textbf{\bibinfo{volume}{49}},
  \bibinfo{pages}{C6} (\bibinfo{year}{1988}),
  \urlprefix\url{http://dx.doi.org/doi/10.1051/jphyscol:1988630}.

\bibitem[{\citenamefont{Bronez}(1988)}]{1988:Bronez}
\bibinfo{author}{\bibfnamefont{T.}~\bibnamefont{Bronez}},
  \bibinfo{journal}{Acoustics, Speech and Signal Processing, IEEE Transactions
  on} \textbf{\bibinfo{volume}{36}}, \bibinfo{pages}{1862}
  (\bibinfo{year}{1988}), ISSN \bibinfo{issn}{0096-3518}.

\bibitem[{\citenamefont{Liu and van Veen}(1992)}]{1992:Liu}
\bibinfo{author}{\bibfnamefont{T.-C.} \bibnamefont{Liu}} \bibnamefont{and}
  \bibinfo{author}{\bibfnamefont{B.}~\bibnamefont{van Veen}},
  \bibinfo{journal}{Signal Processing, IEEE Transactions on}
  \textbf{\bibinfo{volume}{40}}, \bibinfo{pages}{578} (\bibinfo{year}{1992}),
  ISSN \bibinfo{issn}{1053-587X}.

\bibitem[{\citenamefont{Hanssen}(1997)}]{1997:Hanssen}
\bibinfo{author}{\bibfnamefont{A.}~\bibnamefont{Hanssen}},
  \bibinfo{journal}{Signal Process.} \textbf{\bibinfo{volume}{58}},
  \bibinfo{pages}{327} (\bibinfo{year}{1997}), ISSN \bibinfo{issn}{0165-1684}.

\bibitem[{\citenamefont{{Wieczorek} and {Simons}}(2005)}]{2005:Wieczorek}
\bibinfo{author}{\bibfnamefont{M.~A.} \bibnamefont{{Wieczorek}}}
  \bibnamefont{and} \bibinfo{author}{\bibfnamefont{F.~J.}
  \bibnamefont{{Simons}}}, \bibinfo{journal}{Geophysical Journal International}
  \textbf{\bibinfo{volume}{162}}, \bibinfo{pages}{655} (\bibinfo{year}{2005}).

\bibitem[{\citenamefont{{Dahlen} and {Simons}}(2007)}]{2007:Dahlen}
\bibinfo{author}{\bibfnamefont{F.~A.} \bibnamefont{{Dahlen}}} \bibnamefont{and}
  \bibinfo{author}{\bibfnamefont{F.~J.} \bibnamefont{{Simons}}},
  \bibinfo{journal}{ArXiv e-prints} \textbf{\bibinfo{volume}{705}}
  (\bibinfo{year}{2007}), \eprint{0705.3083}.

\bibitem[{\citenamefont{{Simons} et~al.}(2004)\citenamefont{{Simons}, {Dahlen},
  and {Wieczorek}}}]{2004:SDW}
\bibinfo{author}{\bibfnamefont{F.~J.} \bibnamefont{{Simons}}},
  \bibinfo{author}{\bibfnamefont{F.~A.} \bibnamefont{{Dahlen}}},
  \bibnamefont{and} \bibinfo{author}{\bibfnamefont{M.~A.}
  \bibnamefont{{Wieczorek}}}, \bibinfo{journal}{ArXiv Mathematics e-prints}
  (\bibinfo{year}{2004}), \eprint{math/0408424}.

\bibitem[{\citenamefont{{Tegmark}}(1997)}]{1997:Tegmark}
\bibinfo{author}{\bibfnamefont{M.}~\bibnamefont{{Tegmark}}},
  \bibinfo{journal}{\prd} \textbf{\bibinfo{volume}{55}}, \bibinfo{pages}{5895}
  (\bibinfo{year}{1997}), \eprint{arXiv:astro-ph/9611174}.

\bibitem[{\citenamefont{{Oh} et~al.}(1999)\citenamefont{{Oh}, {Spergel}, and
  {Hinshaw}}}]{1999:Oh}
\bibinfo{author}{\bibfnamefont{S.~P.} \bibnamefont{{Oh}}},
  \bibinfo{author}{\bibfnamefont{D.~N.} \bibnamefont{{Spergel}}},
  \bibnamefont{and}
  \bibinfo{author}{\bibfnamefont{G.}~\bibnamefont{{Hinshaw}}},
  \bibinfo{journal}{\apj} \textbf{\bibinfo{volume}{510}}, \bibinfo{pages}{551}
  (\bibinfo{year}{1999}), \eprint{arXiv:astro-ph/9805339}.

\end{thebibliography}

\end{document}